\documentclass[traditabstract]{aa}

\usepackage{graphicx}
\usepackage{txfonts}

\usepackage{natbib}
\bibpunct{(}{)}{;}{a}{}{,}

\def \vv#1{\mathbf{#1}}

\def\pfe#1#2#3{\left(\frac{#1}{#2}\right)^{#3}}
\def \be {\begin{equation}}
\def \ee {\end{equation}}

\def\Deltae {\Delta e}
\def\Deltax {\Delta x}
\def\G {{\cal G}}
\def\m {m}
\def\M {m_0}
\def\p {1}
\def\pp {{}}
\def\c {2}
\def\ij {i}
\def\ji {j}
\def\gr {0}
\def\vpi {\varphi}

\def\cte {Cte}

\def\Kl {K'}
\def\w {\omega}
\def\cA {{\cal A}}
\def\cB {{\cal B}}

\def\crm{\cr\noalign{\medskip}}
\newcommand{\FFrac}[2]{{{\displaystyle\strut#1}\over{\displaystyle\strut#2}}}

\def \llabel#1{\label{#1}}

\begin{document}

\title{Tidal damping of the mutual inclination in hierachical systems}
\titlerunning{Tidal damping of the mutual inclination}

\author{
A.C.M.~Correia\inst{1,2}
\and G.~Bou\'e\inst{2,3}
\and J.~Laskar\inst{2}
\and M.H.M.~Morais\inst{1}
}

 
\institute{Departamento de F\'isica, I3N, Universidade de Aveiro, Campus de
Santiago, 3810-193 Aveiro, Portugal
  \and 
ASD, IMCCE-CNRS UMR8028,
Observatoire de Paris, UPMC,
77 Av. Denfert-Rochereau, 75014 Paris, France  
\and 
Department of Astronomy and Astrophysics, University of Chicago, 5640 S. Ellis Ave, Chicago, IL 95064, USA
}

\date{Received ; accepted To be inserted later}

\abstract{
Hierarchical two-planet systems, in which the inner body's semi-major axis is between 0.1 and 0.5~AU, usually present high eccentricity values, at least for one of the orbits.
As a result of the formation process, one may expect that planetary systems with high eccentricities also have high mutual inclinations.
However, here we show that tidal effects combined with gravitational interactions damp the initial mutual inclination to modest values in timescales that are shorter than the age of the system.
This effect is not a direct consequence of tides on the orbits, but it results from a secular 
forcing of the inner planet's flattening.
We then conclude that these hierarchical planetary systems are unlikely to present very high mutual inclinations, at least as long as the orbits remain outside the Lidov-Kozai libration areas.
The present study can also be extended to systems of binary stars and to planet-satellite systems.
}

   \keywords{celestial mechanics -- planetary systems -- planets and satellites: general}

   \maketitle
%



\section{Introduction}


Today, nearly 100 multi-planet systems have been reported, of which
roughly $1/3$ possess at least one ``moderate close-in planet'', that is, a planet with a semi-major axis between 0.1 and 0.5~AU\footnote{http://exoplanet.eu}.
Planets in this range are supposed to undergo significant tidal interactions, resulting in slowly modified spins and orbits.
However, for the typically assumed dissipation rates for gaseous planets, the spin of moderate close-in planets reaches an equilibrium state in only a few million years, while the orbital evolution can last for the entire age of the system (Gyr timescale).

Among two-planet systems, there is a special class whose semi-major ratio $a_\p / a_\c $ is lower than 0.1, the so-called ``hierarchical systems''. 
This class counts at least 20 members ($1/5$ of all multi-planet systems), and usually  at least one of the planets' orbits is highly eccentric\footnotemark[\value{footnote}].  
During the formation process, the orbital eccentricities can  increase through gravitational scattering \citep[e.g.][]{Nagasawa_etal_2008}, which is  simultaneously responsible for an increase of the orbits' mutual inclination \citep[e.g.][]{Chatterjee_etal_2008}.
Evidently, smaller mass planets, that are as yet undetected, can exist in these systems, but for a semi-major axis larger than 0.1~AU  the orbits usually present high values in the eccentricities, which reduces the stability areas for additional companions.

Hierarchical systems are particularly interesting from a dynamical point of view, because they can be stable with very eccentric and inclined orbits, which is an uncommon behavior.
In particular, they are very interesting when the inner planet is sufficiently close to the star to undergo tidal interactions, since the final outcome of the evolution can be in a configuration that is totally different from the initial one. 
Because the semi-major ratio is small, low-order mean motion resonances cannot occur, which allows us to perform analytical approximations such as averaging the orbits over the mean anomalies.
In addition, tidal effects usually act over very long timescales and therefore
approximate theories also allow one to accelerate the numerical simulations and to
explore the parameter space much more rapidly.

Secular perturbation theories based on series expansions have been used for hierarchical systems.  
For low eccentricity values, the expansion of the perturbation in eccentricity series is very efficient \citep[e.g.][]{Wu_Goldreich_2002}, but this method is not appropriate for orbits that become very eccentric. An expansion in the ratio of the semi-major axis $a_\p / a_\c $ is then  preferred, because exact expressions can be computed for the secular system \citep[e.g.][]{Laskar_Boue_2010}. 
The development to the second order in  $a_\p / a_\c $, called the quadrupole approximation, was used by \citet{Lidov_1961,Lidov_1962} and \citet{Kozai_1962} for the restricted inner problem (the outer orbit is unperturbed).
In this case, the conservation of the normal component of the angular momentum enables the inner orbit to periodically exchange its eccentricity with inclination (the so-called Lidov-Kozai mechanism).
For planar problems, the series expansions in  $a_\p / a_\c $ should be conducted to the octupole
order \citep[e.g.][]{Marchal_1990,
Ford_etal_2000,Laskar_Boue_2010}, because the quadrupole approximation  
fails to reproduce the eccentricity oscillations \citep[e.g.][]{Lee_Peale_2003}. 
Since we do not have any restrictions for the eccentricities or for the mutual inclination, we need to expand the gravitational potential in $a_\p / a_\c $ to the octupole order.

The ultimate stage for tidal evolution is the spin synchronization and orbital circularization \citep[e.g.][]{Correia_2009}.
Indeed, the observed mean eccentricity for planets and binary stars with $ a_\p < 0.1 $~AU is close to zero within the observational limitations \citep[e.g.][]{Pont_etal_2011}.
Although tidal effects modify the spin on a much shorter timescale than they modify the orbit, synchronous rotation can only occur when the eccentricity is very close to zero: the rotation rate tends to be locked with the orbital speed at the periapsis, because tidal effects are stronger when the two bodies are closer to each other.
In addition, if there is a companion body, the eccentricity oscillates \citep[e.g.][]{Mardling_2007, Laskar_etal_2012}, and the rotation rate of the planet shows variations that follow the eccentricity.
This is exactly what is observed for Mercury, whose average orbital eccentricity is around 0.2, and its rotation is captured in a 3/2 spin-orbit resonance \citep{Correia_Laskar_2004,Correia_Laskar_2012}.
Therefore, spin and orbital evolution cannot be dissociated, and some unexpected behavior can be observed, such as a secular increase of the eccentricity \citep[e.g.][]{Correia_etal_2011, Correia_etal_2012}.

In this paper we intend to intensify the study of hierarchical two-planet systems, in which the inner orbit undergoes tidal dissipation.
We present here another counterintuitive behavior, the inclination damping, which is also a consequence of the above-mentioned eccentricity pumping when the two orbits are not coplanar.

\section{Model}

\llabel{TheModel}

We considered a system consisting of a central star of mass $\M$, an inner planet of mass $\m_\p$, and an outer companion of mass $\m_\c$.
We used Jacobi canonical coordinates, with $\vv{r_\p} $ being the position of
$\m_\p$ relative to $\M$, and $ \vv{r_\c} $ the position of $ \m_\c $
relative to the center of mass of $ \m_\p $ and $ \M $.
We assumed that the system is hierarchical, thus $|\vv{r_\p}|  \ll  |\vv{r_\c}|$. 
For simplicity, we express all the angles in the invariable plane of the system, i.e., the plane perpendicular to the total angular momentum
\vskip-.5cm
\be
\vv{H} = \vv{L} + \vv{G}_\p + \vv{G}_\c \ , \llabel{120329c}
\ee
where $\vv{L}$ is the rotational angular momentum of the inner planet, and $\vv{G}_\ij $ the orbital angular momentum of each body.

The inner planet is considered an oblate ellipsoid with gravity field coefficients given by $J_2$, rotating about the axis of maximal inertia (gyroscopic approximation), 
with rotation rate $\omega$, such that \citep[e.g.][]{Lambeck_1988}
\vskip-.5cm
\be
J_2 = k_2 \frac{\omega_\pp^2 R_\pp^3}{3 \G \m_\p} \ .
\llabel{101220a}
\ee
$\G$ is the gravitational constant, $R_\pp$ is the radius of the planet, and
$ k_2 $ is the second Love number for potential (pertaining to a perfectly fluid body).
We furthermore assumed that the obliquity of the planet to the orbital plane is zero ($\varepsilon = 0^\circ$), that is, $\vv{L}$ and $\vv{G}_\p$ are aligned.
Therefore, the angle between the two orbital planes, i.e., the mutual inclination $I$, satisfies the 
relation
\vskip-.5cm
\be
H^2 = (L + G_\p)^2 + G_\c^2 +{2 (L + G_\p) G_\c}\cos I \ .
\llabel{120329d}
\ee

\subsection{Conservative motion}

Because we are interested in the secular behavior of the system, we averaged the equations of motion over the mean anomalies of both orbits.
In the invariable plane, the averaged potential, quadrupole-level for the spin \citep[e.g.][]{Correia_Laskar_2010},  octupole-level for the orbits \citep[e.g.][]{
Ford_etal_2000,Laskar_Boue_2010}, and with general relativity corrections \citep[e.g.][]{Touma_etal_2009} is given by

\vskip-.6cm
\begin{eqnarray}
U & =&  -  \frac{C_0}{(1-e_\p^2)^{1/2}} - \frac{C_1}{(1-e_\p^2)^{3/2}}   \crm & &
- C_2 \frac{(1 + \frac{3}{2} e_\p^2) (1-\frac{3}{2} \sin^2 I)}{(1-e_\c^2)^{3/2}}
- C_2 \frac{ \frac{15}{4} e_\p^2 \sin^2 I}{(1-e_\c^2)^{3/2}} \cos 2 \w_\p \crm & &
+ C_3 \frac{ \cA }{(1-e_\c^2)^{5/2}} e_\p e_\c \cos \vpi \crm & &
+ C_3 \frac{\frac{5}{2} (1-e_\p^2) \cos I \sin^2 I}{(1-e_\c^2)^{5/2}} e_\p e_\c \sin \w_\p \sin \w_\c
\ , \llabel{090514a}
\end{eqnarray}
where 
\be
C_0 = \frac{3 \beta_\p \G^2 (\M + \m_\p)^2}{a_\p^2 c^2} \ , \quad
C_1 = \frac{\G \M  \m_\p J_2 R_\pp^2}{2 a_\p^3}  \llabel{110816a} \ ,
\ee
\be
C_2 = \frac{\G \beta_\p \m_\c  a_\p^2}{4 a_\c^3 }  \ , \quad
C_3 = \frac{15 \G \beta_\p \m_\c a_\p^3}{16 a_\c^4} \frac{(\M-\m_\p)}{\M + \m_\p} \llabel{110816c} \ ,
\ee
\be
\cA=  1 + \frac{3}{4} e_\p^2 - \frac{5}{4} \cB  \sin^2 I  \llabel{120402a} \ ,
\ee
\be
\cB  =  1+ \frac{5}{2} e_\p^2 - \frac{7}{2} e_\p^2 \cos 2 \w_\p  \llabel{120329a} \ ,
\ee
and $\vpi$ is the angle between the directions of the periastrons:
\be
\cos \vpi = - \cos \w_\p \cos \w_\c - \sin \w_\p \sin \w_\c \cos I  \llabel{120329a} \ .
\ee
$a_\ij$ is the semi-major axis (that can also be expressed using the mean motion $n_\ij $),
$e_\ij$ is the eccentricity, and $\w_\ij $ is the argument of the periastron.
We also have $ \beta_\p = \M m_\p / (\M + \m_\p) $,  $ \beta_\c = (\M + \m_\p) \m_\c / (\M + \m_\p + \m_\c) $, 
$ G_\ij = \beta_\ij n_\ij a_\ij^2 (1-e_\ij^2)^{1/2}$, and
$ L = \xi m_\p R_\pp^2 \omega_\pp $, where $ \xi $ is the normalized moment of inertia.

The contributions to the orbits are easily obtained using the Lagrange planetary equations \citep[e.g.][]{Murray_Dermott_1999}:
\be
\dot e_\ij = \frac{\sqrt{1-e_\ij^2}}{\beta_\ij n_\ij a_\ij^2 e_\ij} \frac{\partial U}{\partial \w_\ij} \ , \quad
\dot \w_\ij = - \frac{\sqrt{1-e_\ij^2}}{\beta_\ij n_\ij a_\ij^2 e_\ij} \frac{\partial U}{\partial e_\ij} \llabel{110816d} \ .
\ee
In addition, since the variations in $e_\p$, $e_\c$ and $I$ are related by the conservation of the total angular momentum (Eq.\,\ref{120329d}), we have  
\begin{eqnarray}
\frac{\partial \cos I}{\partial G_\p} &= -\left[ \FFrac{1}{G_\c} + \FFrac{1}{L + G_\p} \cos I \right] \ , \crm
\frac{\partial \cos I}{\partial G_\c} &= -\left[ \FFrac{1}{L + G_\p} + \FFrac{1}{G_\c} \cos I \right] \ . 
\end{eqnarray}
As we  assumed  that $L / G_\p \sim (R_\pp / a_\p)^2 \ll 1$, if we neglect first-order terms in $L/G_1$, 
we simply have  ($\ij \ne \ji = 1, 2$):

\be
\frac{\partial \cos I}{\partial e_\ij} =  \frac{G_\ij \, e_\ij}{1-e_\ij^2} \frac{\partial \cos I}{\partial G_\ij} \approx \frac{e_\ij}{1-e_\ij^2} \left( \frac{G_\ij}{G_\ji} + \cos I \right) \ . \llabel{120329e}
\ee

Thus,
\begin{eqnarray}
\dot e_\p & = & \nu_{21} \frac{ \frac{5}{2}  (1-e_\p^2)^{1/2}  \sin^2 I}{(1-e_\c^2)^{3/2}} e_\p \sin 2 \w_\p \crm 
&-&  \nu_{31}  \frac{ \cA (1-e_\p^2)^{1/2}}{(1-e_\c^2)^{5/2}} e_\c \sin \vpi_\p \crm
&-&   \nu_{31}  \frac{ \frac{35}{4}  e_\p^2 \sin 2 \w_\p \sin^2 I}{(1-e_\p^2)^{-1/2} (1-e_\c^2)^{5/2}} e_\c \cos \vpi \crm
&+&   \nu_{31} \frac{\frac{5}{2} (1-e_\p^2)^{3/2}  \cos I \sin^2 I}{(1-e_\c^2)^{5/2}} e_\c \cos \w_\p \sin \w_\c 
\ , \llabel{110816h}
\end{eqnarray}
\begin{eqnarray}
\dot e_\c & = &  \nu_{32} \frac{ \cA }{(1-e_\c^2)^{2}} e_\p \sin \vpi_\c \crm
&+& \nu_{32} \frac{\frac{5}{2} (1-e_\p^2) \cos I \sin^2 I}{(1-e_\c^2)^{2}} e_\p \sin \w_\p \cos \w_\c \ , \llabel{110816i} 
\end{eqnarray}
and

\begin{eqnarray}
\dot \w_\p &=& \frac{\nu_\gr}{ (1-e_\p^2)} + \frac{\nu_1 \, x_\pp^2}{(1-e_\p^2)^2} \crm 
&+& \nu_{21}  \frac{2 (1-e_\p^2) + \frac{5}{2} (e_\p^2 - \sin^2 I) (1 - \cos 2 \w_\p)}{(1-e_\p^2)^{1/2}  (1-e_\c^2)^{3/2}} \crm 
&+& \nu_{22} \frac{(1 + \frac{3}{2} e_\p^2 - \frac{5}{2} e_\p^2 \cos 2 \w_\p ) \cos I}{(1-e_\c^2)^{2}} \crm 
&-& \nu_{31} \frac{ \cA + \frac{3}{2} e_\p^2 - \frac{5}{4} (5 -7 \cos 2 \w_\p )e_\p^2 \sin^2 I}{e_\p (1-e_\p^2)^{-1/2} (1-e_\c^2)^{5/2}} e_\c \cos \vpi \crm 
&-& \nu_{31} \frac{ \frac{5}{2} \cB  e_\p^2 \cos^2 I}{e_\p (1-e_\p^2)^{1/2} (1-e_\c^2)^{5/2}}  e_\c \cos \vpi \crm
&+& \nu_{31} \frac{\cA e_\p^2 + \frac{5}{2} (1-e_\p^2) (2 e_\p^2 - \sin^2 I)}{e_\p (1-e_\p^2)^{1/2} (1-e_\c^2)^{5/2}}  e_\c \cos I \sin \w_\p \sin \w_\c \crm
&-& \nu_{32} \frac{\frac{5}{2} \cB \cos I}{(1-e_\c^2)^{3}} e_\p e_\c \cos \vpi \crm 
&+& \nu_{32} \frac{\cA - \frac{5}{2} (1-e_\p^2) (1 - 3 \cos^2 I)}{(1-e_\c^2)^{3}} e_\p e_\c \sin \w_\p \sin \w_\c
 \ , \llabel{110819a1}
\end{eqnarray}
\begin{eqnarray}
\dot \w_\c &=& \nu_{21} \frac{(1 + \frac{3}{2} e_\p^2 -  \frac{5}{2} e_\p^2 \cos 2 \w_\p) \cos I}{(1-e_\p^2)^{1/2} (1-e_\c^2)^{3/2}} \crm 
&+& \nu_{22} \frac{1 + \frac{3}{2} e_\p^2 + (1-\frac{5}{2} \sin^2 I) (1 + \frac{3}{2} e_\p^2 -  \frac{5}{2} e_\p^2 \cos 2 \w_\p)}{(1-e_\c^2)^{2}} \crm 
&-& \nu_{31} \frac{ \frac{5}{2} \cB  \cos I}{(1-e_\p^2)^{1/2}(1-e_\c^2)^{5/2}} e_\p e_\c  \cos \vpi \crm
&+& \nu_{31} \frac{ \cA - \frac{5}{2} (1-e_\p^2) (1 - 3 \cos^2 I)}{(1-e_\p^2)^{1/2}(1-e_\c^2)^{5/2}} e_\p e_\c \sin \w_\p \sin \w_\c \crm
&-& \nu_{32} \frac{ \cA  (1+4e_\c^2) +  \frac{5}{2} \cB e_\c^2 \cos^2 I}{e_\c (1-e_\c^2)^{3}} e_\p \cos \vpi \crm
&+& \nu_{32} \frac{\cA e_\c^2 + 5 (1-e_\p^2) (e_\c^2 - \frac{1}{2}(1+ 7e_\c^2) \sin^2 I)}{e_\c (1-e_\c^2)^{3}} \quad \times \crm
&& \quad \quad \quad \quad \quad \quad \quad \quad \quad \quad \quad
 e_\p   \cos I \sin \w_\p \sin \w_\c 
\ , \llabel{110819a2} 
\end{eqnarray}
where $ x_\pp = \omega_\pp / n_\p $, the constant frequencies
\be
\nu_\gr = 3 n_\p \pfe{n_\p a_\p}{c}{2} \ , \llabel{110817f}
\ee
\be
\nu_1 = n_\p \frac{k_2}{2} \frac{\M + \m_\p}{\m_\p} \pfe{R_\pp}{a_\p}{5} \ , \llabel{110817a}
\ee
\be
\nu_{21} =  n_\p \frac{3}{4} \frac{\m_\c}{\M + \m_\p} \pfe{a_\p}{a_\c}{3} \ , \llabel{110817b}
\ee
\be
\nu_{22} =  n_\c \frac{3}{4} \frac{\M \m_\p}{(\M +\m_\p)^2} \pfe{a_\p}{a_\c}{2}  \ , \llabel{110817c}
\ee
\be
\nu_{31} =  n_\p \frac{15}{16} \frac{\m_\c}{\M+\m_\p} \frac{\M-\m_\p}{\M+\m_\p} \pfe{a_\p}{a_\c}{4} \ , \llabel{110817d}
\ee
\be
\nu_{32} =  n_\c \frac{15}{16} \frac{\M \m_\p}{(\M+\m_\p)^2} \frac{\M-\m_\p}{\M+\m_\p} \pfe{a_\p}{a_\c}{3} \ , \llabel{110817e}
\ee
and
\be
\sin \vpi_\p = - \frac{\partial (\cos \vpi)}{\partial \w_\p} = - \sin \w_\p \cos \w_\c + \cos \w_\p \sin \w_\c \cos I  \llabel{120329a1} \ ,
\ee
\be
\sin \vpi_\c = \frac{\partial (\cos \vpi)}{\partial \w_\c}  = \cos \w_\p \sin \w_\c - \sin \w_\p \cos \w_\c \cos I  \llabel{120329a2} \ .
\ee
When $ I = 0^\circ $, we have $ \vpi = \vpi_\p = \vpi_\c = \w_\c - \w_\p $. 
Note also that the longitude of the node does not appear in the equations of motion (Eqs.\,\ref{110816h}$-$\ref{110819a2}) because we used the invariable plane as the reference plane (Eq.\,\ref{090514a}), and thus $\Delta \Omega = 180^\circ$.


\begin{table*}
\begin{center}
\caption{Hierarchical two-planet non-resonant systems with $0.1 < a_\p < 0.5 $~AU and 
$ a_\p / a_\c < 0.1 $.
 \llabel{table1}}
\begin{tabular}{lcccccccccccc}
\hline
\hline
Star & Age& $\M$ & $\m_\p$ & $\m_\c$ & $a_\p$ & $e_\p$ & $a_\c$ & $e_\c$ & $\vpi^* $ & $R_\pp $  & $I_\mathrm{max}$ & Ref. \\
(name) & (Gyr) & ($M_\odot$) & ($M_{J}$)  & ($M_{J}$)  & (AU) &   & (AU) &   & (deg) & ($R_\mathrm{Jup}$)  & (deg) &  \\
\hline
HD\,190360 & 7.8 & 0.96 & .057  & 1.50  & 0.128 & 0.01  & 3.92 & 0.36 & 219. & 0.54 & $\sim 40^\star$ &  [1] \\
HD\,38529   & 3.3 & 1.48 & 1.13  & 17.6 & 0.131 & 0.25  & 3.70 & 0.36 & 269. & 1.13 & $\sim 40^\star$ &  [2] \\
HD\,11964   & 9.6 & 1.12   & .079  & 0.62 & 0.229 & 0.30  & 3.16 & 0.04 & 307. & 0.60 & $\sim 40$ &  [3,4] \\
HD\,147018 &  6.4 & 0.93   & 2.12  & 6.56 & 0.239 & 0.47  & 1.92 & 0.13 & 251. & 1.25 & $\sim 40$ &  [5] \\
HD\,168443 &  9.8 & 0.99   & 7.66  & 17.2 & 0.293 & 0.53  & 2.84 & 0.21 & 252. & 1.51 & $\sim 50$ &  [6] \\
HD\,74156   & 3.7 & 1.24 & 1.88  & 8.03  & 0.294 & 0.64  & 3.40 & 0.43 & 266. & 1.23 & $\sim 20$ &  [7,8,9] \\
HD\,163607 & 8.6 & 1.09   & 0.77  & 2.29 & 0.360 & 0.73  & 2.42 & 0.12 & 186. & 1.05 & $\sim 30$ &  [10] \\
\hline
\end{tabular}
\end{center}
Notes: All masses $m_\ij$ correspond to minimum values ($I_\ij = 90^\circ$), except for HD\,38529, which has $I_\ij = 48^\circ$; $\vpi^* = \w_\c^* -\w_\p^* $; $R_\pp $ was estimated using Eq.(\ref{120529b}); The maximal inclination $I_\mathrm{max}$ was estimated using $k_2 \Delta t = 100 $\,s, $(^\star)$ for HD\,190360 and HD\,38529 starting with  $a_\p = 0.2$~AU and $e_\p = 0.25$.
References: 
[1] \citet{Vogt_etal_2005}; 
[2] \citet{Benedict_etal_2010}; 
[3] \citet{Butler_etal_2006}; [4] \citet{Wright_etal_2009}; 
[5] \citet{Segransan_etal_2010}; 
[6] \citet{Pilyavsky_etal_2011};
[7] \citet{Naef_etal_2004}; 
[8] \citet{Bean_etal_2008};
[9] \citet{Meschiari_etal_2011}; 
[10] \citet{Giguere_etal_2012}.
\end{table*}

\subsection{Tidal effects}

\llabel{TidalEffects}

In our model, we additionally considered tidal dissipation raised by the central star on the inner planet.
The dissipation of the mechanical energy of tides in the planet's interior is responsible for a time delay $\Delta t_\pp$ between the initial perturbation and the maximal deformation. 
Because the rheology of planets is poorly known, the exact dependence of $\Delta t_\pp$ on the tidal frequency is unknown.
Several models exist \citep[for a review see][]{Correia_etal_2003,Efroimsky_Williams_2009}, but for simplicity we adopted a model with constant $\Delta t_\pp$, 
which can be made linear \citep{Singer_1968,Mignard_1979}.
The contributions to the equations of motion are given by \citep[e.g.][]{Correia_2009, Correia_etal_2011}
\be
\frac{\dot \omega_\pp}{n_\p} = - K_\pp
\left( f_1(e_\p) x_\pp - f_2(e_\p)  \right) \ , \llabel{090515a}
\ee
\be
\frac{\dot a_\p}{a_\p} = 2 \Kl_\pp \,
\left( f_2(e_\p) x - f_3(e_\p) \right) \ , \llabel{090515b}
\ee
\be\dot e_\p = 9 \Kl_\pp \left( \frac{11}{18} f_4(e_\p) x - f_5(e_\p) \right) e_\p \ ,
\llabel{090515c}
\ee
\be
\dot I = - \frac{\Kl_\pp }{2} \frac{f_1(e_\p)}{(1-e_\p^2)^{1/2}}  x_\pp  \sin \varepsilon \, = 0  \ ,
\llabel{120529a}
\ee
where
\be
K_\pp =  n_\p \frac{3 k_2}{\xi Q} \frac{\M \beta_\p}{\m_\p^2} \pfe{R_\pp}{a_\p}{3}  \ ,
\ee
\be
\Kl_\pp = \frac{K_\pp}{1/\xi}
\frac{\m_\p}{\beta_\p} \pfe{R_\pp}{a_\p}{2} \llabel{090514m} \ , 
\ee
\be Q_\pp^{-1} \equiv n_\p \Delta t_\pp \ , \llabel{120704a} \ee
and
\be f_1(e) = \frac{1 + 3e^2 + 3e^4/8}{(1-e^2)^{9/2}}  \ , \ee 
\be f_2(e) = \frac{1 + 15e^2/2 + 45e^4/8 + 5e^6/16}{(1-e^2)^{6}} \ , \ee
\be f_3(e) = \frac{1 + 31e^2/2 + 255e^4/8 + 185e^6/16 + 25e^8/64}{(1-e^2)^{15/2}} \ , \ee
\be f_4(e) = \frac{1 + 3e^2/2 + e^4/8}{(1-e^2)^5} \ , \ee
\be f_5(e) = \frac{1 + 15e^2/4 + 15e^4/8 + 5e^6/64}{(1-e^2)^{13/2}} \ . \ee

We neglected the effect of tides over the argument of the periastron, as well as the flattening of the central star. Their effect is only to add a small supplementary frequency to $\dot \w_\p$, similar to the contributions from the general relativity 
\citep[for a complete model see][]{Correia_etal_2011}.
Expression (\ref{120529a}) for the inclination is zero, because we assumed the obliquity of the planet to be zero ($\varepsilon = 0^\circ)$.

Under the effect of tides alone,
the equilibrium rotation rate, obtained when $ \dot \omega_\pp = 0 $, is attained for (Eq.\,\ref{090515a})
\be
\frac{\omega_\pp}{n_\p} = f(e_\p) = \frac{f_2(e_\p)}{f_1(e_\p)} = 1 + 6 e_\p^2 + {\cal O}(e_\p^4)
 \ . \llabel{090520a}
\ee
Usually, $ \Kl_\pp \ll K_\pp $, so tidal effects modify the rotation rate much faster than the orbit. It is thus tempting to replace the equilibrium rotation in expressions (\ref{090515b}) and (\ref{090515c}).
With this simplification, one always obtains negative contributions for $ \dot a_\p $ and $ \dot e_\p $ \citep{Correia_2009}, 
\be
\frac{\dot a_\p}{a_\p} = - 7 K_\pp' \, f_6(e_\p) e_\p^2 \, < 0
\ , \llabel{090522a}
\ee
\be
\dot e_\p = - \frac{7}{2} K_\pp' f_6(e_\p) (1-e_\p^2) e_\p \, < 0 
\ , \llabel{090522b}
\ee
with 
\be
f_6 (e) = \frac{1 + \frac{45}{14} e^2 + 8e^4 + \frac{685}{224} e^6 + \frac{255}{448} e^8 + \frac{25}{1792}e^{10}}{(1-e^2)^{15/2}  (1 + 3e^2 + 3e^4/8)} \ . \llabel{090527a}
\ee
Thus, the semi-major axis and the eccentricity can only decrease until the orbit of the planet becomes circular.
However, planet-planet interactions can produce eccentricity
oscillations with a period shorter than, or comparable to, the damping timescale of the
spin. In that case, expression (\ref{090520a}) is not satisfied
and multi-planetary systems may show non-intuitive eccentricity evolutions,
such as eccentricity pumping of the inner orbit \citep{Correia_etal_2012}.

\begin{figure}[ht!]
\centering
\includegraphics[width=8.5cm]{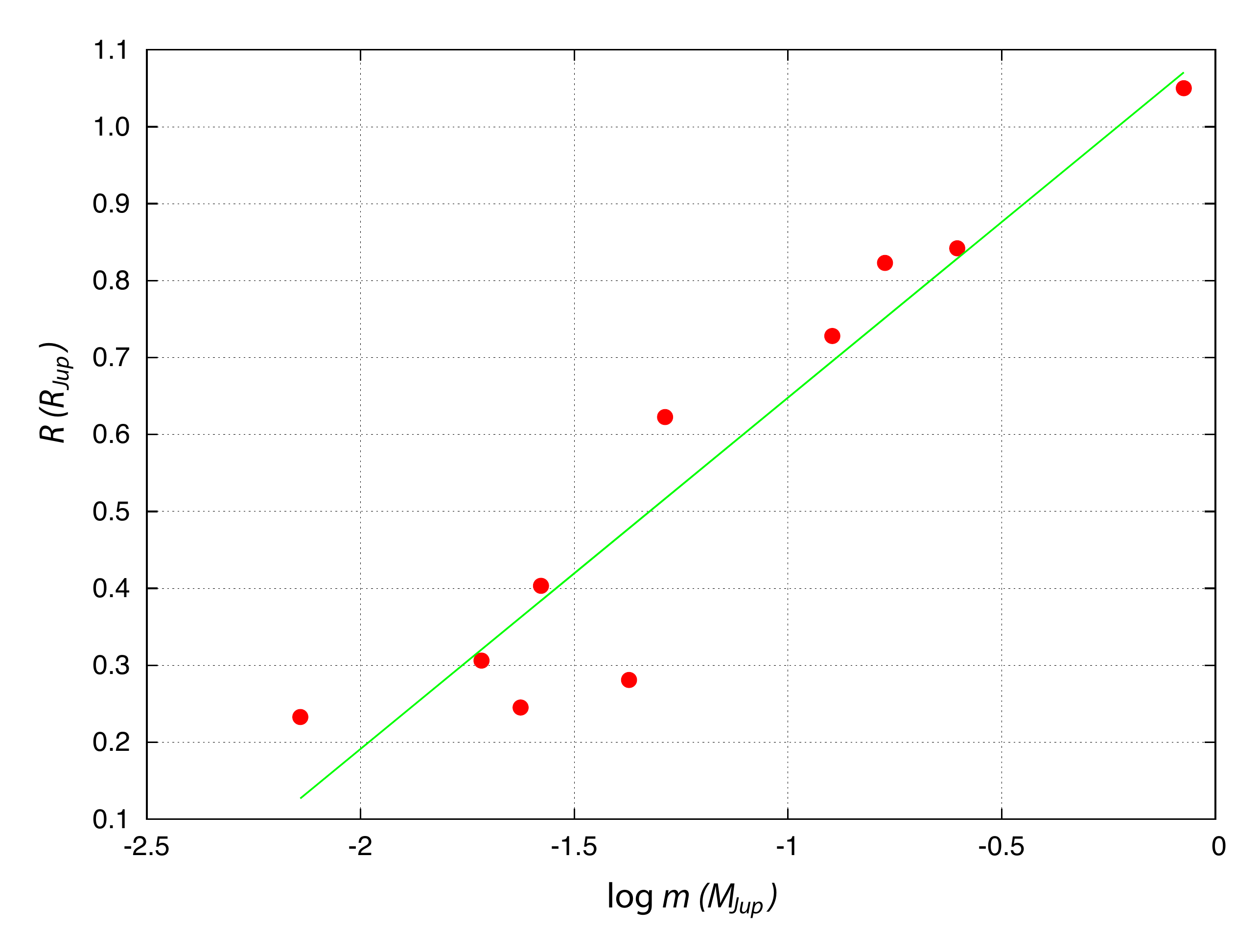}
\caption{Radius versus the mass of the planet. We plot all known close-in planets in the range $ 0.1 < a_\p < 0.5 $~AU,  for which the radius was determined by the transiting method. We observe that the radius decreases with the mass in a relatively regular way.  \llabel{fig1}}
\end{figure}

\begin{figure*}[ht!]
\includegraphics[width=18cm]{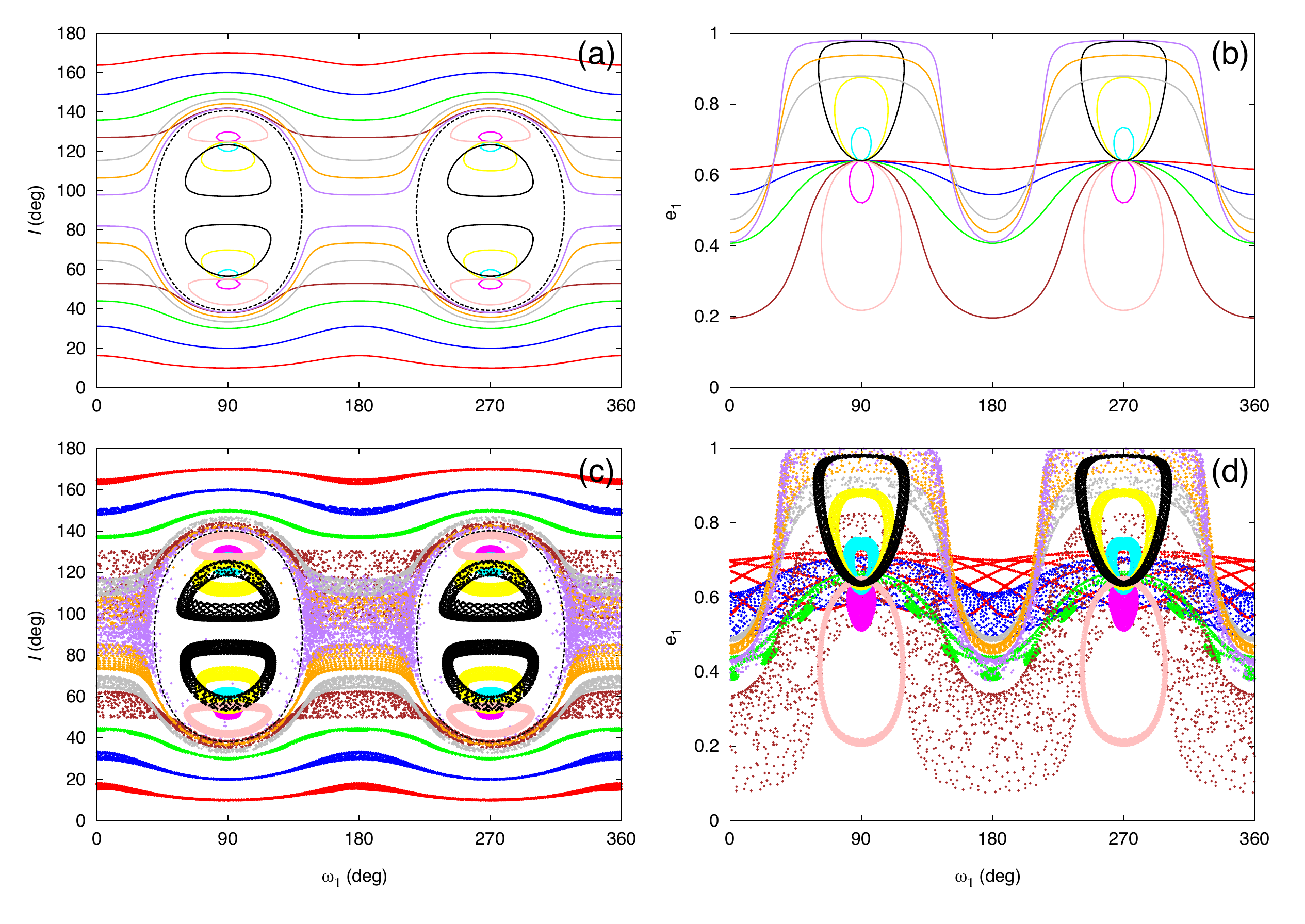}
\caption{Possible secular trajectories for the HD\,74156 system seen in the $(\w_\p,I)$ plane ({\it left}), and in the $(\w_\p, e_\p)$ plane ({\it right}). We show the trajectories using the quadrupolar approximation ({\it top}), corresponding to the level curves of constant energy, and the octupolar approximation ({\it bottom}). The dashed black curves in the  $(\w_\p,I)$ plane correspond to the separatrix between the circulation and libration zones of $ \w_\p$. The colors are preserved in all pictures, each one corresponding to a given value of the total angular momentum of the system (determined by different values of the initial mutual inclination). All trajectories are compatible to the present knowledge for this system (Table\,\ref{tabD}).   \llabel{fig2}}
\end{figure*}

\begin{table}
\begin{center}
\caption{Stability analysis of the HD\,74156 system for different sets ($\w_\p, I$) of initial conditions (Fig. \ref{fig2}).
Chaotic diffusion is present whenever $D > 10^{-6}$ (in bold). 
 \llabel{tabD}}
\begin{tabular}{lccrr}
\hline
\hline
trajectory & $\w_\p$ & $I_0$ & $2 \pi/g$ & $\log D$  \\
(color) & (deg) & (deg)  & (kyr)  &   \\
\hline
\multicolumn{5}{c}{\it circulation} \\
\hline
red & 90 & 10  & 13.657  & -9.41 \\
blue & 90 & 20  & 15.065 & -9.12  \\
green & 90 & 30  & 18.554  & -7.66 \\
brown & 90 & 38  & 20.877 & {\bf -4.78}  \\
gray & 30 & 60  & 16.992 & {\bf -5.90}  \\
orange & 30 & 70  & 17.773  & {\bf -5.33} \\
purple & 30 & 80  & 20.289 & {\bf -1.60} \\
\hline
\multicolumn{5}{c}{\it libration} \\
\hline
pink & 90 & 42  & 15.482 & -9.08  \\
mangenta & 90 & 50  & 9.918  & -7.20 \\
cyan & 90 & 60  & 7.703 & -8.50  \\
yellow & 90 & 70  & 6.804  & -9.99 \\
black & 90 & 83  & 6.330  & -7.60 \\
\hline
\end{tabular}
\end{center}
\end{table}

\begin{figure}[ht!]
\centering
\includegraphics[width=8.5cm]{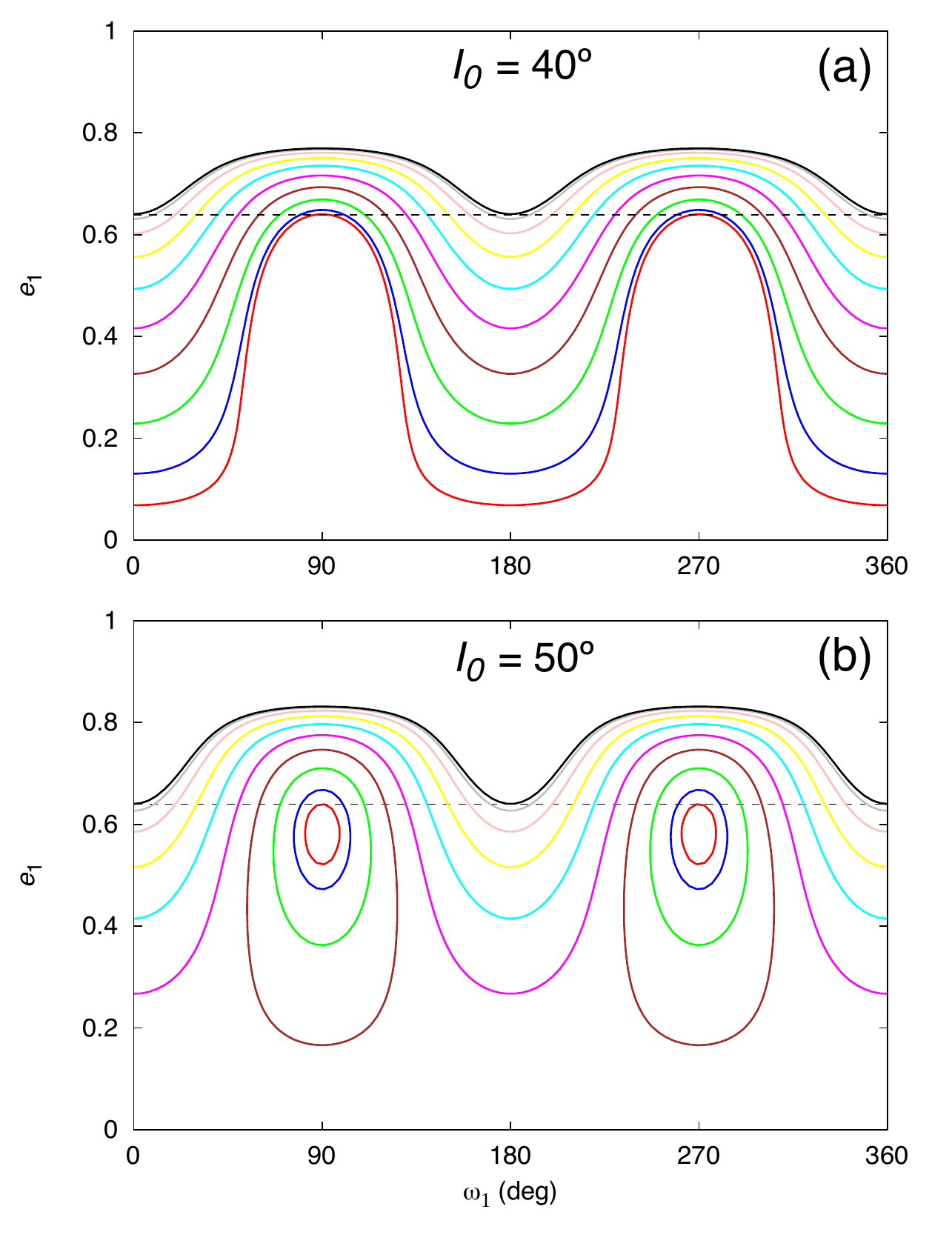}
\caption{Possible secular trajectories for the HD\,74156 system seen in the $(\w_\p, e_\p)$ plane for two different values of the initial mutual inclination, $I_0 = 40^\circ$ ({\it top}), and $I_0 = 50^\circ$ ({\it bottom}). We show the trajectories using the quadrupolar approximation, corresponding to the level curves of constant energy.  Each one corresponds to an initial value of the argument of the periastron $ \w_\p$, ranging from $0^\circ$ ({\it black}) to $90^\circ$ ({\it red}) with a step of $10^\circ$. The dashed black curves corresponds to the observed eccentricity of the planet, that is, it gives the initial condition for $ \w_\p$. All trajectories are compatible with the current knowledge for this system (Table\,\ref{table1}). \llabel{fig3}}
\end{figure}

\begin{figure}[ht!]
\includegraphics[width=8.5cm]{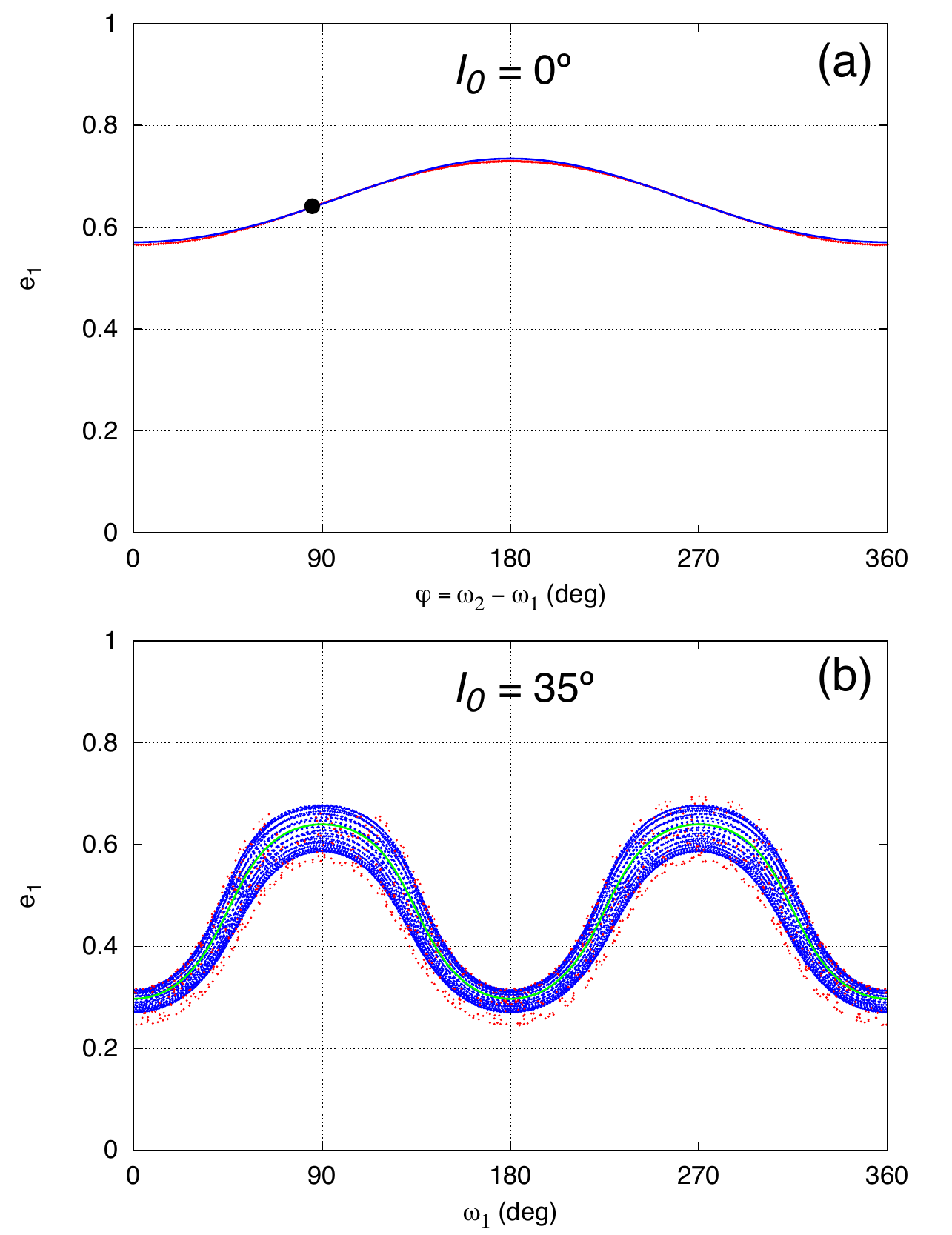}
\caption{Possible secular trajectories for the HD\,74156 system seen in the $(\vpi, e_\p)$ plane for $I=0^\circ$ ({\it top}) and in the $(\w_\p, e_\p)$ plane for $I=35^\circ$ ({\it bottom}).
All trajectories are compatible with the current knowledge for this system (Table\,\ref{table1}).
We show the trajectories using the octupolar approximation (blue) and direct numerical simulations (red). In (a) the blue path corresponds to the level curves of constant energy for coplanar orbits. The dot marks the present position of the planet. In (b) we additionally show the trajectories using the quadrupolar approximation (green), which corresponds to the level curves of constant energy.  \llabel{fig4}}
\end{figure}

\section{Application to exoplanets}

In the following sections we apply the model described in Section~\ref{TheModel} to different configurations of hierarchical two-planet systems.
To observe the damping effect of the mutual inclination, the spin of the planet must be fully damped, but not its orbit, that is, $ \Kl_\pp \ll K_\pp $ (see Appendix~\ref{InclinationDamping}).
In addition, the damping timescale of the spin should be of the same order as the period of the eccentricity oscillations,  $  K_\pp \sim g $ (Eq.\,\ref{110902b}).
This is valid for gaseous planets roughly within $0.1 < a_\p < 0.5 $~AU, which we call ``moderate close-in planets''.
In Table~\ref{table1} we list all hierarchical systems known to date whose inner orbit satisfies the above condition.
We focus our analysis on the HD\,74156 system, but all the main results are easily extended to the remaining planetary systems.

\subsection{Radius of close-in exoplanets}

According to expression (\ref{110902b}), the $\nu_1$ frequency (Eq.\,\ref{110817a}) is a key parameter for the observation of the eccentricity pumping of the inner orbit and consequent damping of the inclination (by means of $ g_x $).
The minimum masses and the semi-major axis are relatively well determined from the observations, so the largest incertitudes in $\nu_1$ come from the Love number $k_2$, and particularly from the radius of the planet, which appears as a power of 5.
Therefore, a correct estimate of the planetary radius is necessary to observe some effect on the inclination.

Since the radius of the planet is correlated with its mass, one solution is to adopt a constant value for the density, $\rho_\pp$, and then compute the radius simply as $R_\pp^3 = 3 m_\p / (4 \pi \rho_\pp)$.
However, by applying this strategy to the two largest planets of the solar system, Jupiter and Saturn, we immediately see that it can give very distinct results.
The density of a planet depends on many factors, such as the age of the system, the initial composition of the accretion disk, or where the planet formed in the disk.
Any theoretical estimation of the radius is then subject to large incertitude, and only direct observations can give reliable values.

We used an empirical expression based on the previously observed radius of close-in planets in the range $ 0.1 < a_\p < 0.5 $~AU. 
We found ten planets in this range\footnotemark[\value{footnote}] whose radius were determined by the transiting method (Fig.\,\ref{fig1}).
We observe that the radius decreases with the mass in a relatively regular way, therefore we performed a linear regression of the observational data:
\be
R_\pp / R_\mathrm{Jup}  = 0.46 \, \log_{} \left( m_\pp / M_\mathrm{Jup} \right) + 1.10
\ . \llabel{120529b}
\ee
This expression also agrees well with the solar system data, giving $1.1  \, R_\mathrm{Jup}$ for Jupiter and $ 1.0 \, R_\mathrm{Sat}$ for Saturn.

\subsection{Initial conditions uncertainty}

\llabel{uncertainty}

Assuming that the observational values of the minimum masses, semi-major axis, and eccentricities of the planetary systems listed in Table~\ref{table1} are relatively well determined, we can use them as a starting point to study these systems.

A striking observation is that the eccentricity of at least one of the planets can be very high.
Because hierarchical systems exhibit high eccentricity values, the mutual inclinations can also be very high \citep[e.g.][]{Chatterjee_etal_2008}.
However, currently their mutual inclinations are unknown, not only because we are unable to determine the inclination with respect to the plane of the sky, $I_i$ (and hence the true masses), but also because we are unable to determine the longitude of the nodes in the plane of the sky, $\Omega_i$:
\be
\cos I = \cos I_\p \cos I_\c + \sin I_\p \sin I_\c \cos (\Omega_\c - \Omega_\p) \ .
\llabel{120627a}
\ee
The only partial exception is the system HD\,38529, for which $I_\c \approx 48^\circ$, estimated using astrometric measurements from the {\it Hubble Space Telescope} \citep{Benedict_etal_2010}.
The planetary masses $m_\p$ and $m_\c$ given in Table~\ref{table1} correspond to the minimum masses (assuming $I_\p = I_\c = 90^\circ $), except for HD\,38529, where the masses were estimated using $I_\p = I_\c = 48^\circ $.

According to expression (\ref{120329d}), without knowing the mutual inclination of these systems it is impossible to determine the total angular momentum $H$, which is a constant of the motion.
Even if we assume that the system is coplanar and prograde ($\cos I = 1$), the total angular momentum is undetermined because the true planetary masses appearing in the expressions of $G_\p$ and $G_\c$ are also unknown (except for HD\,38529).
As a consequence, the present dynamics of these systems can be considerably different, depending on the true $H$ value.

Moreover, although the argument of the periastron of these planets is known in the plane of the sky (angle $\vpi^* = \w_\p^* - \w_\c^*$ in Table\,\ref{table1}), the arguments of the periastron in the invariable plane of the system ($ \w_\p$ and $ \w_\c$) are also unknown because they depend on $I_i$ and $\Omega_i$ \citep[e.g.][]{Giuppone_etal_2012}:
\be
\cos (\w_\p - \w_\p^*) = \frac{\cos I_\c - \cos I_\p \cos I}{\sin I_\p \sin I} \ ,
\llabel{120629a}
\ee
where $ \w_\ij^*$ is the argument of the periastron in the plane of the sky.
For $\w_\c$ we have an identical relation, the only difference is that $\w_\p$ is measured from the ascending node, while $\w_\c$ is measured from the descending node.

Therefore, apart from the semi-major axes, eccentricities and minimum masses, currently there are few constraints on the remaining orbital parameters of hierarchical systems.
Because we are only concerned with the tidal evolution of the mutual inclination, we adopted the masses listed in Table\,\ref{table1} for the numerical simulations.
The only free parameters are then the mutual inclination and the arguments of the periastron, whose uncertainty are related to the lack of knowledge of the longitude of the node in the plane of the sky  $(\Omega_\c-\Omega_\p)$.

Using the quadrupolar approximation for gravitational interactions, the potential energy (Eq.\,\ref{090514a}) is independent of the $ \w_\c$, and therefore $e_\c$ is constant (Eq.\,\ref{110816d}).
The dynamics of the system is then fully described by the couples $(I, \w_\p)$ or  $(e_\p, \w_\p)$.
Adopting the minimum masses (Table\,\ref{table1}), we show in Figure~\ref{fig2}(a,b) some possibilities for the HD\,74156 system for different mutual inclinations (corresponding to different $H$ values).
For nearly coplanar systems ($I < 20^\circ $), only small variations are observed for $e_\p$ and $I$.
However, as the inclination increases, the dynamics of the system is considerably perturbed by the presence of Lidov-Kozai cycles \citep{Lidov_1961,Lidov_1962,Kozai_1962}.
In this regime, we can observe significant exchanges between the inclination and the eccentricity of the inner orbit. 
In some cases the eccentricity can reach values much higher than today, and thus enhance the tidal dissipation.

\begin{figure*}[ht!]
\includegraphics[width=18cm]{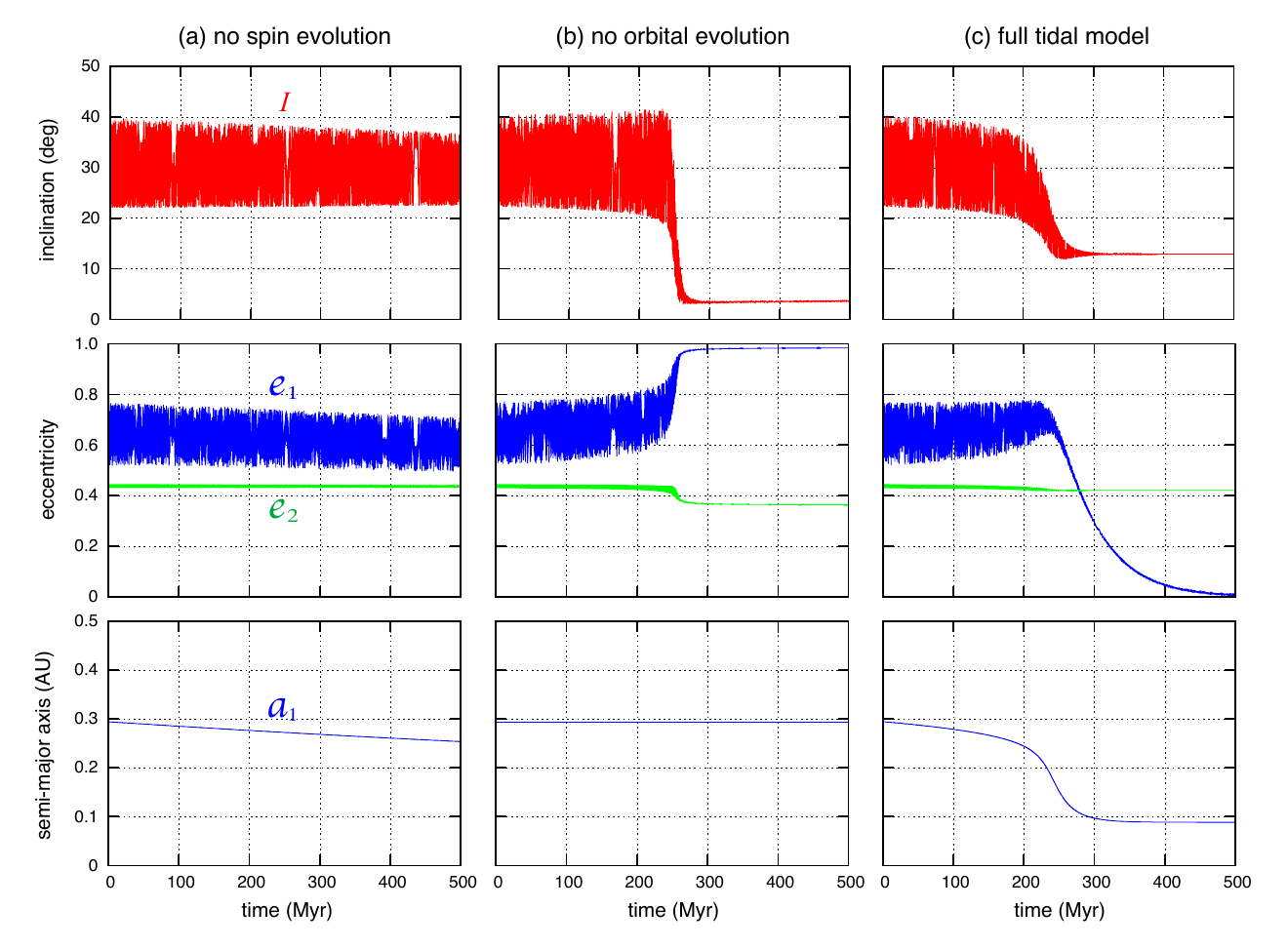}
\caption{Long-term evolution of the HD\,74156 system for different tidal models with $I_0 = 40^\circ$: only dissipation in the orbit is considered (a),  only dissipation in the spin is considered (b), full model (c). We plot the mutual inclination $I$ ({\it top}), the eccentricities $e_\p$ (blue) and $e_\c$ (green) ({\it middle}), and the semi-major axis $a_\p$ ({\it bottom}).   \llabel{fig5}}
\end{figure*}

\begin{figure*}[ht!]
\includegraphics[width=18cm]{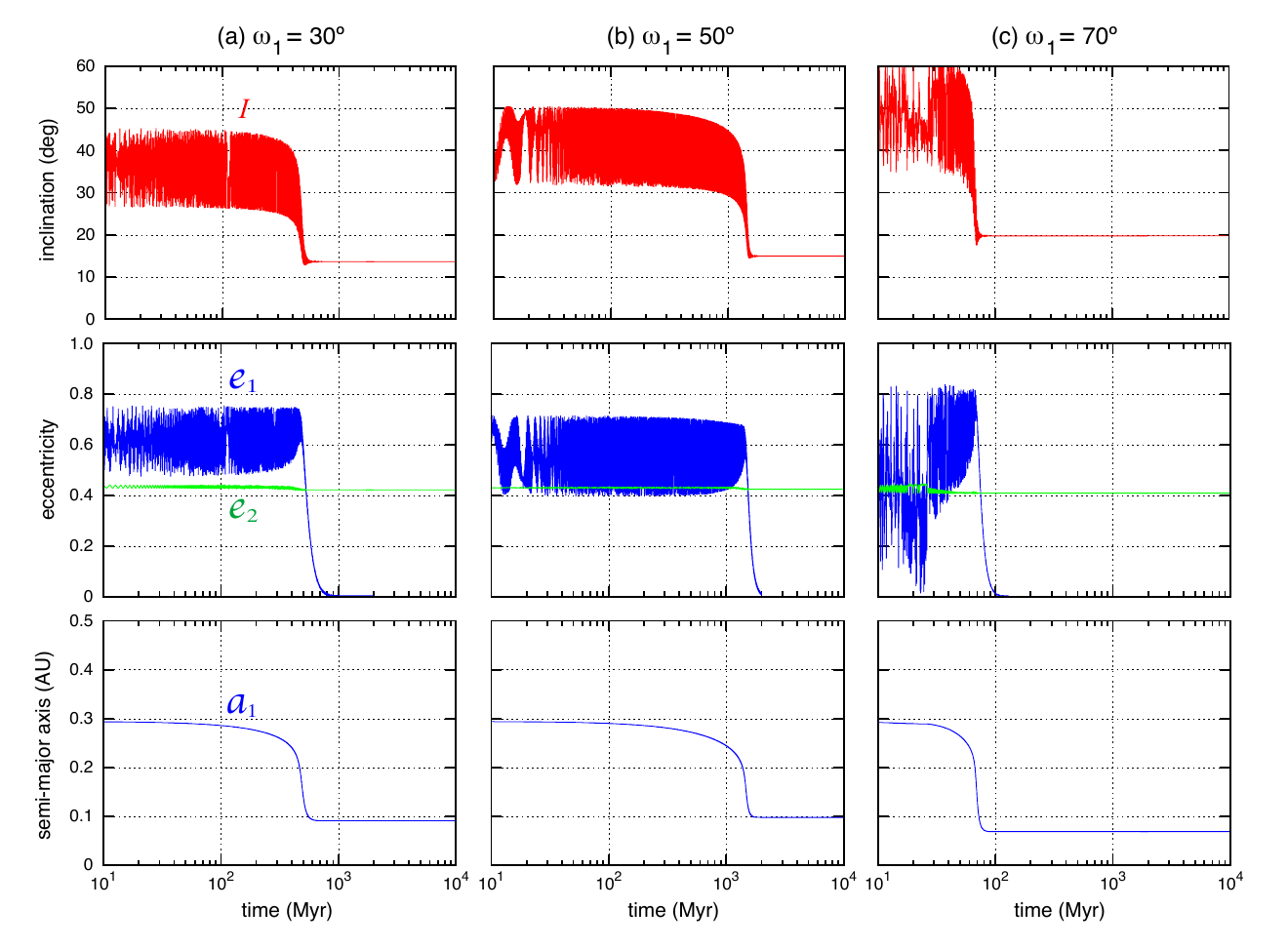}
\caption{Long-term evolution of the HD\,74156 system with $I_0 = 40^\circ$ for different values of the argument of the periastron, $ \w_{\p} = 30^\circ$ (a),  $50^\circ$ (b), and $ 70^\circ$ (c). We plot the mutual inclination $I$ ({\it top}), the eccentricities $e_\p$ (blue) and $e_\c$ (green) ({\it middle}), and the semi-major axis $a_\p$ ({\it bottom}).   \llabel{fig6}}
\end{figure*}

\begin{figure*}[ht!]
\includegraphics[width=18cm]{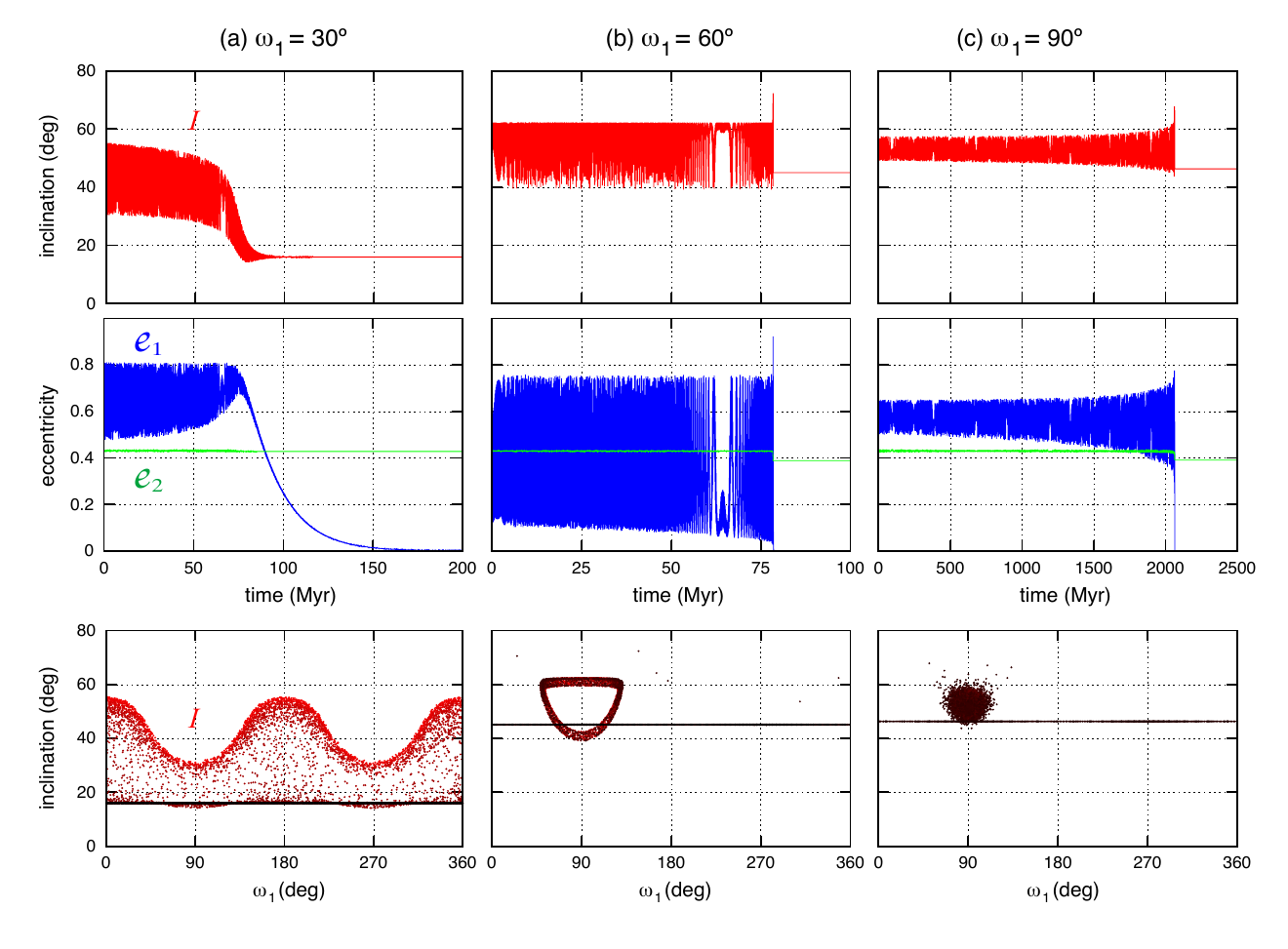}
\caption{Long-term evolution of the HD\,74156 system with $I_0 = 50^\circ$ for different values of the argument of the periastron, $ \w_{\p} = 30^\circ$ (a),  $60^\circ$ (b), and $ 90^\circ$ (c). As a function of time, we plot the mutual inclination $I$ ({\it top}) and the eccentricities $e_\p$ (blue) and $e_\c$ (green) ({\it middle}). As a function of $ \w_\p$, we plot again the mutual inclination, where the color of each dot becomes darker with time ({\it bottom}).  \llabel{fig7}}
\end{figure*}

Figure~\ref{fig2}(a,b) is different from the standard Lidov-Kozai diagrams that show level curves of the quadrupole Hamiltonian at fixed values  $G_1 \cos I$, because for HD\,74156 the initial eccentricity of the inner planet is already fixed at $e_\p = 0.64$ (Table\,\ref{table1}). 
Instead, we  followed the procedure in \citet{Giuppone_etal_2012} and show the trajectories for different values of the initial mutual inclination $I$, i.e., we varied the total angular momentum $H$. This explains why librating orbits in the Lidov-Kozai regime do not encircle a Lidov-Kozai equilibrium point (which occurs at the current eccentricity value).  

The impact of the initial uncertainty on the $ \w_\p$ value is shown in Figure~\ref{fig3}. 
Depending on this value, the observed eccentricity can correspond to a maximum or to a minimum for an identical total angular momentum $H$ (Fig.\,\ref{fig3}a). 
Moreover, two trajectories may be in circulation or in libration (Fig.\,\ref{fig3}b).
It is therefore very important to completely explore the phase space of the initial conditions $(I, \w_\p)$ to capture the global dynamics of hierarchical two-planet systems.

\subsection{Octupole contribution}

So far, we restricted our analysis to the quadrupolar gravitational interactions, because they are mainly responsible for the inclination variations.
However, for the hierarchical systems listed in Table\,\ref{table1}, the range of semi-major ratios is $ 0.03 < a_\p / a_\c < 0.1 $, meaning that octupolar interactions cannot be neglected. 
Indeed, although the impact of octupolar terms on the eccentricity variations is weaker than that of the quadrupole terms, octupolar interactions are strong enough to produce secular drifts when combined with tidal effects \citep{Correia_etal_2012}.

In the planar prograde case ($\cos I = 1$), the potential energy (Eq.\,\ref{090514a}) only depends on $\vpi = \w_\c - \w_\p $, and therefore $I$ is constant (Eq.\,\ref{120329e}).
The dynamics of the system is then fully described by the couples $(\vpi, e_\p)$ or  $(\vpi, e_\c)$.
Adopting the minimum masses, the angle $\vpi^*$ in the plane of the sky listed in Table\,\ref{table1} corresponds to the angle $(\w_\c - \w_\p + 180^\circ)$ in the invariant plane of the system, because for $I_\p = I_\c = 90^\circ$ we have $ \w_\p = \w_\p^* + 90^\circ $ and $ \w_\c = \w_\c^* - 90^\circ $ (Eq.\,\ref{120629a}).
In Figure~\ref{fig4}(a) we show the expected eccentricity variations for the 
HD\,74156 system in this unique situation for which the system is fully characterized.

With increasing mutual inclination, we are left with four free parameters ($e_\p, \w_\p, e_\c, \w_\c$), and it becomes impossible to capture the dynamics of the systems in a two-dimensional plot.
Nevertheless, we can perform numerical simulations of the equations of motion (Eqs.\,\ref{110816h}$-$\ref{110819a2}) to understand how the octupolar terms modify the quadrupolar approximation (Fig.\,\ref{fig2}).
We observe that the main effect is to add some diffusion around the quadrupolar trajectories.
The diffusion is more pronounced for orbits in circulation around the separatrix ($35^\circ < I < 145^\circ$).
The stability of the orbits can be measured with a frequency analysis \citep{Laskar_1990,Laskar_1993PD}.
We determined the precession frequency $g$ and $g'$ of the argument of the pericenter $\w_\p$ over two consecutive time intervals of length $ T = 5 $\,Myr.
In Table\,\ref{tabD} we compute the difference $ D = | g - g' | / g $, which is a measure of the chaotic diffusion of the trajectories \citep{Correia_etal_2005, Couetdic_etal_2010}.
It should be close to zero for a regular solution, and values with $D > 10^{-6}$ correspond to chaotic motion.
This is observed for all trajectories in circulation close to the separatrix.

In Figure~\ref{fig4}(b) we plot simultaneously the eccentricity evolution obtained with the two approximations for $ I_0 = 35^\circ $, and compare it with direct numerical simulations.
We conclude that 1) the quadrupolar approximation correctly describes the average dynamics in inclined hierarchical systems; 2) the octupolar approximation is essential to derive a more realistic behavior and obtain results similar to direct numerical simulations.

\section{Tidal evolution}


We now include the effect of tides described in section~\ref{TidalEffects} to the conservative equations of the motion (Eqs.\,\ref{110816h}$-$\ref{110819a2}), and perform some numerical simulations.
In all simulations we adopt for the innermost planet $ \xi_\pp = 1/5 $, $ k_2 = 1/2 $, and a dissipation time lag $ \Delta t_\pp = 200 $\,s.
For HD\,74156 this dissipation is equivalent to $ k_2 / Q_\pp \approx 1.4 \times 10^{-4}$ (Eq.\,\ref{120704a}), which is comparable to the value $ k_2 / Q_\pp = (2.3 \pm 0.7) \times 10^{-4}$ estimated for Saturn \citep[][]{Lainey_etal_2012}.
In addition, we always set $ \w_\c = 180^\circ $ and $ 2 \pi / \omega_\pp = 50 $\,day. 
The impact of the initial $ \w_\c$ value can be obtained by adjusting a different value for $ \w_\p$.
Similarly, the initial rotation rate is not a critical initial parameter, since tidal effects quickly bring the rotation near to the equilibrium value (Eq.\,\ref{090520a}).

\subsection{Effect of the spin}

In Figure~\ref{fig5} we show some examples for the evolution of the HD\,74156 planetary system, in three different situations.
The radius of the inner planet is estimated to be $ R_\pp =  1.23 \, R_{Jup} $ (Eq.\,\ref{120529b}), and we initially assume $ I_0 = 40^\circ $ and $ \w_\p = 0^\circ $.

In a first experiment, we only consider tidal effects on the orbit (Eqs.\,\ref{090522a},\,\ref{090522b}), as it is often done in previous studies.
That is, since the rotation of close-in planets evolves very fast, we assume that the spin is locked in its equilibrium position (Eq.\,\ref{090520a}).
We observe that the eccentricity and the semi-major axis of the inner orbit slowly decrease, while the mutual inclination and the eccentricity of the outer orbit only oscillate around a constant mean value (Fig.\,\ref{fig5}a).
Therefore, we conclude that in this case the only effect of tides is to circularize the inner orbit in a timescale longer than the age of the system, a well-know result in the literature \citep[e.g.][]{Correia_Laskar_2010B}.

In a second experiment, we neglect the effects on the orbit and we only consider the effect on the rotation (Eq.\,\ref{090515a}). 
This situation corresponds to the opposite of the previous one, and it is not realistic, but it allows to highlight the importance of not neglecting the rotation rate evolution.
Indeed, although there is no direct dissipative contribution to the eccentricity or to the inclination, we observe that these two parameters undergo significant variations (Fig.\,\ref{fig5}b), the eccentricity of the inner planet rising almost up to 1.
The pumping effect on the eccentricity due to the spin excitation was reported in the planar case by \citet{Correia_etal_2012}, for which only octupolar terms are important.
In the non-planar case, the pumping effect it is even more pronounced, since quadrupole order terms additionally contribute.
In addition, because the angular momentum is mainly exchanged between the inner planet eccentricity and the inclination, while the first increases, the second decreases.
In Appendix~\ref{InclinationDamping} we provide the full explanation for this effect in the frame of the quadrupolar approximation.
We also observe that the eccentricity of the outer planet is slightly damped during this process, because of the octupole order interactions \citep{Correia_etal_2012}.

Finally, since orbital and spin evolution cannot be dissociated, we integrate the full set of equations for the tidal evolution (Eqs.\,\ref{090515a}$-$\ref{120529a}) (Fig.\,\ref{fig5}c).
We observe that the initial behavior of the system is identical to the situation without dissipation on the orbit (Fig.\,\ref{fig5}b), with a significant damping of the mutual inclination.
However, as the eccentricity increases, the inner planet comes closer to the star at periastron, and tidal effects on the orbit become stronger.
As a consequence, the semi-major axis decreases and the damping effect on the eccentricity (Eq.\,\ref{090515c}) overrides the pumping drift (Eq.\,\ref{110902b}).
At this point, the inclination damping is less efficient, and it ceases when the pumping drift is completely gone.
The system ultimately evolves into a circular orbit as usual, but in a considerable much shorter timescale and it is left with a final lower mutual inclination ($\sim 15^\circ$).

\subsection{Effect of the argument of periastron}

In Section~\ref{uncertainty} we have seen that the initial choice of the argument of the periastron of the inner planet, $ \w_\p$, plays an important role on how the present eccentricity is changing (Fig.\,\ref{fig3}).
Similarly, it also changes the initial trend of the inclination: for increasing eccentricity $e_\p$ the inclination decreases, and vice-versa.
In the example from previous section (Fig.\,\ref{fig5}c), we used $ \w_\p = 0^\circ$, that is, we assumed that the observed value of the eccentricity ($e_\p = 0.64$) is a minimum, and the inclination $I = 40^\circ$ a maximum (Fig.\,\ref{fig3}a).

In order to test the impact of the initial argument of the periastron in the HD\,74156 system, in Figure~\ref{fig6} we plot its evolution for different initial $ \w_\p$ values.
We observe similar behavior as before for all situations, the only significant difference being the evolution timescale.
Until $ \w_\p < 50^\circ $ this timescale increases, because the present eccentricity is no longer a minimum value.
As a consequence, the average value of the eccentricity oscillations is shifted down, and tidal dissipation is less effective, since at the periastron the inner planet is farther from the star.

In the quadrupolar approximation, the behavior described above should be maintained up to $ \w_\p < 90^\circ $, for which the observed eccentricity is a maximum and the inclination a minimum (Fig.\,\ref{fig3}a).
However, the fact that the initial inclination increases when the eccentricity decreases has a strong implication when including octupolar terms (Fig.\,\ref{fig2}): for high inclinations the trajectories are closer to the separatrix, which results in a higher oscillation of the eccentricity.
Thus, for $ \w_\p > 60^\circ$ we observe that the evolution timescale is reduced again, since the inner orbit eccentricity is allowed to reach much higher values than those predicted by the quadrupolar approximation.

\begin{figure}
\includegraphics[width=8.5cm]{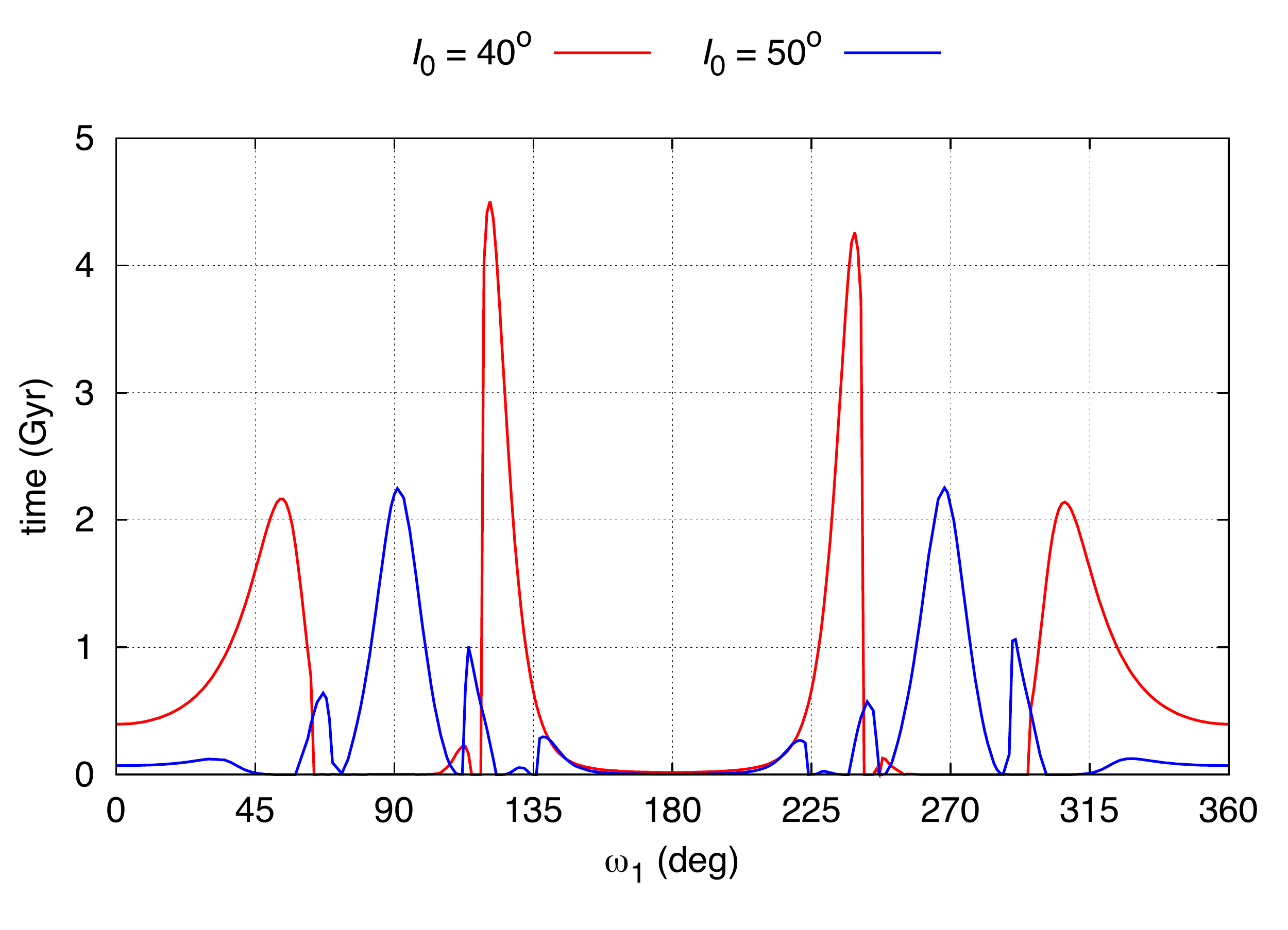}
\caption{Circularization time ($e_\p < 0.01$) for the HD\,74156  system with $I_0 = 40^\circ $ (red)  and $I_0 = 50^\circ $ (blue)  using different values of the initial argument of the pericenter. 
For $I_0 = 40^\circ $ the timescale decreases around the Lidov-Kozai equilibria, because the system is the circulation regime.
For $I_0 = 50^\circ $, the system is in libration, but it is most likely destroyed after it crosses the separatrix.
 \llabel{fig6bis}}
\end{figure}

Up to now, we have been considering an initial mutual inclination  $I_0 = 40^\circ$.
For $I_0\la  40^\circ$, there is only one dynamical regime for the HD\,74256 planetary system, consisting of trajectories in circulation around the Lidov-Kozai equilibira (Fig.\,\ref{fig3}a).
However, for higher values of the initial  inclination we can also observe the libration regime (Fig.\,\ref{fig3}b). 
In Figure~\ref{fig7} we show some numerical simulations with initial $I_0 = 50^\circ$ using different values for the argument of the periastron.

For  $ \w_\p = 30^\circ$ the inner planet is still in circulation (Fig.\,\ref{fig7}a), so we observe identical behavior for the eccentricity and inclination as in the case with $I_0 = 40^\circ$ (Fig.\,\ref{fig6}).
At the bottom of Figure~\ref{fig7} we plot the inclination as a function of the argument of the periastron ($I, \w_\p$).
We plot a dot each $10^5$\,yr and its color becomes darker with time.
There we can clearly see that the planet is always in circulation, and that the amplitude of the oscillations is damped with time.

In the remaining two situations shown in Figure~\ref{fig7},  the inner planet is in libration around the Lidov-Kozai equilibrium located at $ \w_\p = 90^\circ$.
The effect of tides is to slowly increase the amplitude of both the eccentricity and inclination.
As a consequence, the orbit of the planet will cross the separatrix of the libration zone and start to circulate as in the previous case.

For  $ \w_\p = 60^\circ$ (Fig.\,\ref{fig7}b) the planet already starts close to the separatrix, so the initial oscillations are higher and it takes only about 75\,Myr to cross it.
For  $ \w_\p = 90^\circ$ (Fig.\,\ref{fig7}c) the planet is placed close to the Lidov-Kozai equilibrium, so it takes much longer to reach the separatrix.
In both situations, just after the transition of dynamical regime, the eccentricity reaches a very high value close to unity.
Therefore, tidal effects with the central star become very strong and the final evolution is rapid: the semi-major axis decreases, and the inner orbit becomes circular.
However, in a more realistic simulation where we integrate the full equations of motion and take into account the bodies dimensions, the planet most likely collides with the star.
In both situations the system is either destroyed, or its configuration completely modified from the initial situation.

In order to better understand the variation of the evolution timescale with the initial choice of the argument of the pericenter, in Figure~\ref{fig6bis} we plot the circularization time ($ e_\p < 0.01 $) as a function of $ \w_\p$.
The circularization time is more or less equivalent to the inclination damping time, since the final stages of the evolution are very fast.
For $I_0 = 40^\circ $ the timescale decreases around the Lidov-Kozai equilibria, because the system is the circulation regime.
For $I_0 = 50^\circ $, the system is in libration, but it is most likely destroyed after it crosses the separatrix.

\citet{Giuppone_etal_2012} also studied the evolution of planets inside the circulation zone of Lidov-Kozai equilibriums.
They performed some numerical simulations using the quadrupolar approximation and damping of the inner orbit eccentricity due to the presence of a primordial disk.
They concluded that the planet stays in libration and migrates into the Lidov-Kozai equilibrium position, which is exactly the contrary that we observed here.
Since the eccentricity of the inner orbit is also damped in our model, these results appear somehow contradictory.
Therefore, we performed one simulation where the eccentricity is damped, but the semi-major axis is held constant.
In this unrealistic situation, one observe that the planet migrates into the equilibrium like in \citet{Giuppone_etal_2012}.
As a consequence, it seems that there is no inconsistency between the two models, but it becomes clear that the semi-major axis evolution plays an important role in destabilizing the Lidov-Kozai equilibria.
It appears that it cannot be neglected in future studies on the migration of the initial orbits as in \citet{Giuppone_etal_2012}.

\subsection{Constraints for the mutual inclination}

\llabel{constraints}

In previous sections, we saw that in mutually inclined hierarchical two-planet systems there is a significant increase in the eccentricity of the inner planet's orbit.
As a result, the tidal dissipation is enhanced when the planet is at the periastron, and the system evolves faster into an equilibrium configuration.

In Figure~\ref{fig8} we show the evolution of the HD\,74156 system for three different values of the initial inclination $I_0 = 15^\circ$, $30^\circ$ and $45^\circ$.
As expected, when we increase the mutual inclination, the evolution timescale decreases.
For $ I_0 = 15^\circ $ the eccentricity and the semi-major take more than 10~Gyr to be completely damped, while for $I_0 = 45^\circ$ the system is fully evolved only after 100\,Myr.
Moreover, for $I_0 = 15^\circ$ there is almost no effect on the mutual inclination, we only observe some amplitude damping when the eccentricity is decreased to low values, because gravitational perturbations no longer force the inclination.
On the contrary, for $I_0 = 30^\circ$, the pumping effect on the eccentricity is already present, and hence we observe a significant reduction of the final mutual inclination.

\begin{figure*}[ht!]
\includegraphics[width=18cm]{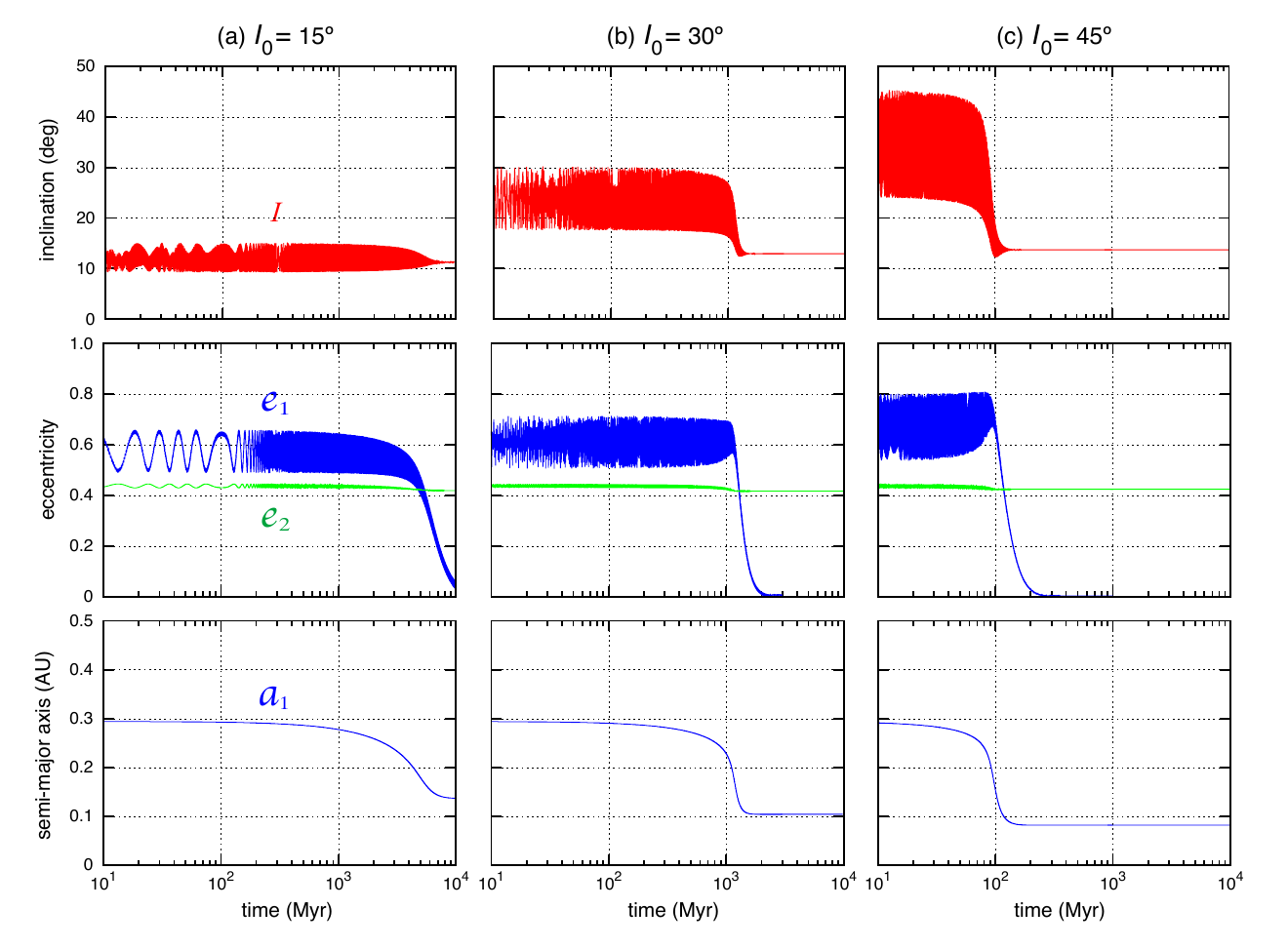}
\caption{Long-term evolution of the HD\,74156 system for different values of the initial inclination, $I_0 = 15^\circ$ (a),  $30^\circ$ (b), and $ 45^\circ$ (c). We plot the mutual inclination $I$ ({\it top}), the eccentricities $e_\p$ (blue) and $e_\c$ (green) ({\it middle}), and the semi-major axis $a_\p$ ({\it bottom}).   \llabel{fig8}}
\end{figure*}

In Figure~\ref{fig9} we show the same kind of evolution as before, but for an initial retrograde orbit with initial inclinations $I_0 = 165^\circ$, $150^\circ$ and $135^\circ$.
In this case, the evolution of the system does not differ much from the prograde situation, the only significant difference is that the inclination is damped to high values close to $180^\circ$ (coplanar system with a retrograde orbit).
For $I_0 = 165^\circ$ the inclination is more or less conserved and the eccentricity is damped over 10~Gyr, while for lower values of the initial inclination, the inclination is damped and the system  evolves in much shorter timescales.

\begin{figure*}[ht!]
\includegraphics[width=18cm]{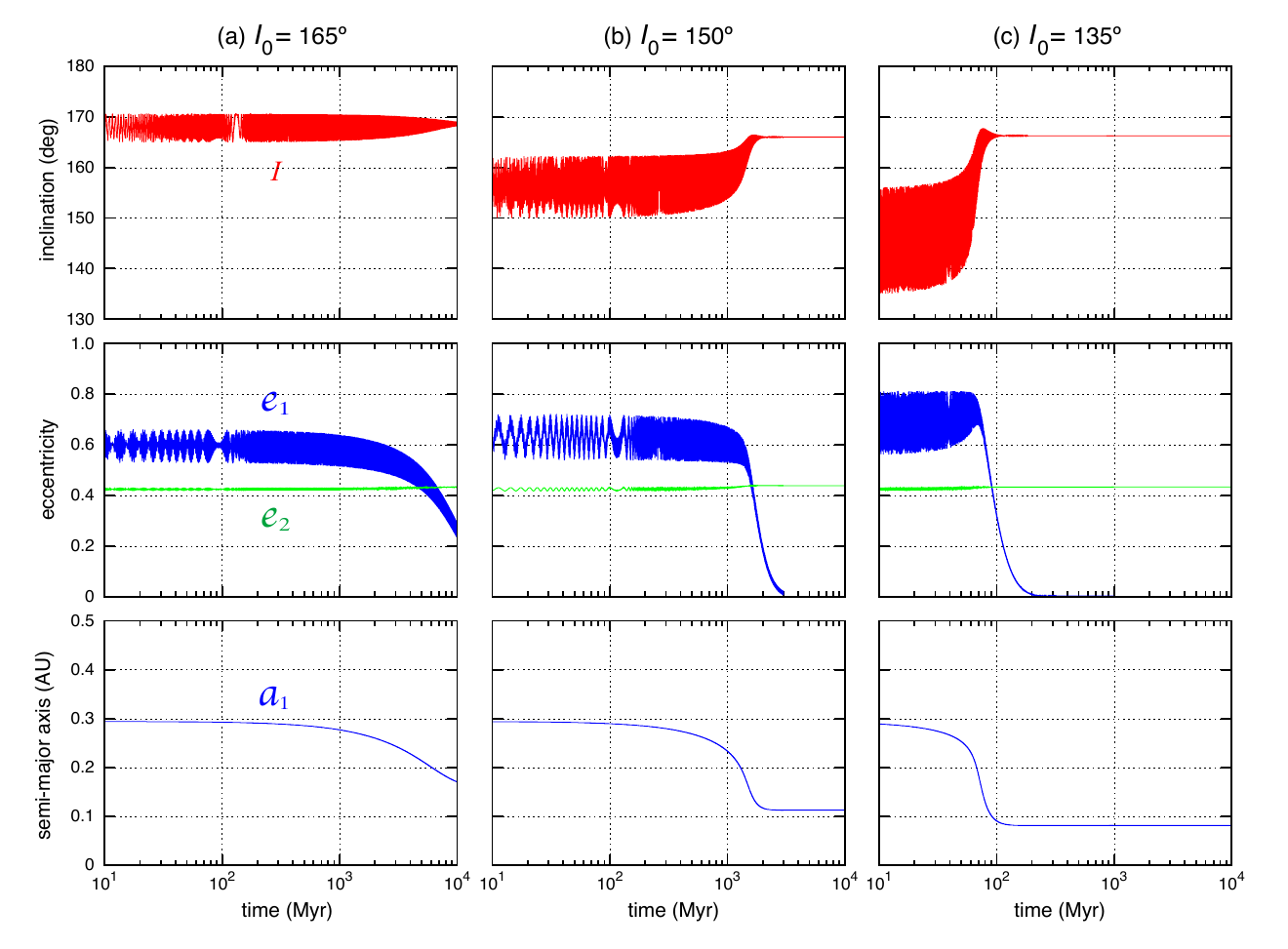}
\caption{Long-term evolution of the HD\,74156 system for different values of the initial inclination, $I_0 = 165^\circ$ (a),  $150^\circ$ (b), and $135^\circ$ (c). We plot the mutual inclination $I$ ({\it top}), the eccentricities $e_\p$ (blue) and $e_\c$ (green) ({\it middle}), and the semi-major axis $a_\p$ ({\it bottom}).   \llabel{fig9}}
\end{figure*}

From the observation of Figures~\ref{fig8} and~\ref{fig9} we then conclude that mutual inclinations closer to $90^\circ$ speed up the final evolution of the system. 
Since most hierarchical systems listed in Table\,\ref{table1} (except HD\,190360 and HD\,38529) still present substantial values for the inner-orbit's eccentricity, we then expect that their mutual inclinations are not extremely high.
For all those systems we run several numerical simulations starting with the present initial conditions from Table\,\ref{table1}, adopting $ k_2 \Delta t = 100 $\,s, and different initial values for $I$ and $ \w_\p$. 
All trajectories that circularize the inner-orbit in less than $\sim 10$\,Gyr 
can then be ruled out, while those not showing significant modifications can be retained as possible representations of the real system.
Therefore, we are able to set some constraints for the maximal mutual inclination of each system, whose limitations are listed in Table\,\ref{table1}.
In Figure~\ref{fig10} we show the outcome of these simulations for the HD\,74156 system, which corresponds to a summary of the more detailed evolutions shown in previous Figures.
Orbits with $20^\circ < I < 150^\circ$ circularize the system in less than 10~Gyr, so they can be discarded.

When we run the same kind of simulations for the HD\,38529, we observe that the eccentricity is damped very quickly, even for coplanar orbits.
One possibility is that we overestimated the dissipation. 
However, even if we adopt $ k_2 \Delta t = 10 $\,s, that is, one order of magnitude lower than for the remaining planets, the system still circularizes in a timescale shorter than the age of the system. 
Another possibility is to suppose that 
the inner planet semi-major axis was higher in the past.
This hypothesis can also be extended to the HD\,190360 system, for which the inner orbit is already circularized, but it may have had a higher eccentricity value in the past.
In order to test this scenario, 
for all planetary systems in Table\,\ref{table1}, we run several numerical simulations for different initial values for $I$ and $ \w_\p$, keeping $ k_2 \Delta t = 100 $\,s, but adopting $a_\p = 0.2 $~AU and $ e_\p = 0.25 $ as initial values, instead of the current values.
In Figure~\ref{fig12} we plot an example for the HD\,190360 system.
By modifying the initial conditions, we are able to reproduce the present observations.
Note that the HD\,190360 system is older than the HD\,38529 one (Table\,\ref{table1}), so both systems may have undergone an identical evolution, but they are observed at different stages. 
The initial semi-major axis could have been higher, providing that the inner orbit eccentricity was also higher (for instance $a_\p = 0.25 $~AU and $ e_\p = 0.4 $).

\begin{figure}[ht!]
\includegraphics[width=8.5cm]{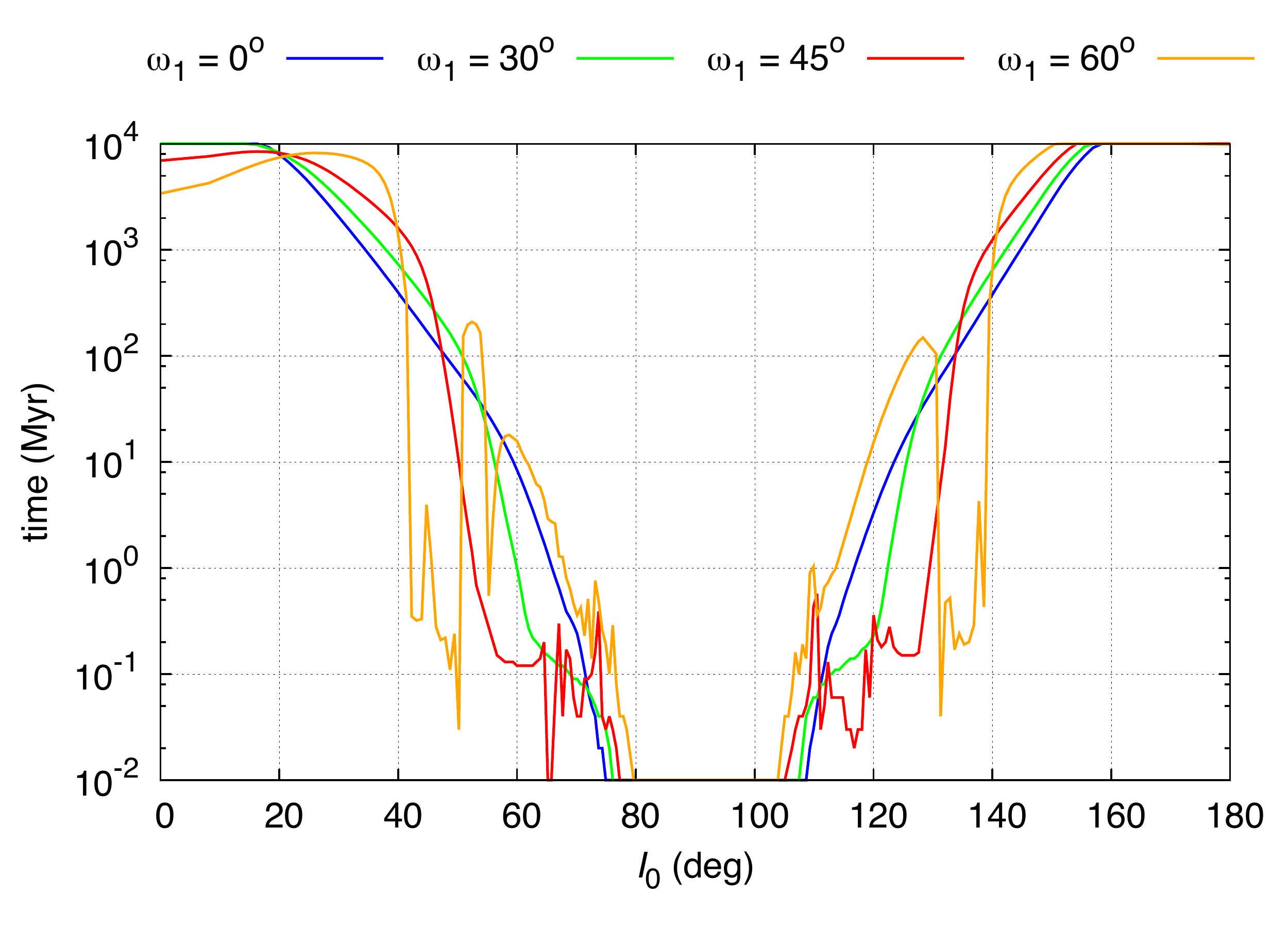}
\caption{Circularization time ($e_\p < 0.01$) for the HD\,74156  system using different values of the initial mutual inclination and argument of the pericenter. Since the estimated age of the system is 
several Gyr (Table\,\ref{table1}), we can rule out mutual inclinations within $ 20^\circ < I < 150^\circ$.  \llabel{fig10}}
\end{figure}

\begin{figure}[ht!]
\includegraphics[width=8.5cm]{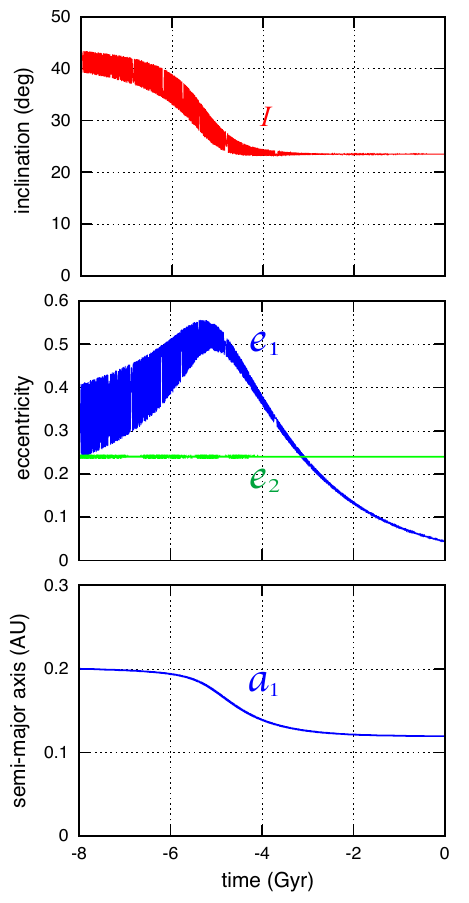}
\caption{Possible past evolution of the HD\,190360 system with $I_0 = 43^\circ$, $a_\p = 0.2 $~AU, and $e_\p = 0.25$. We plot the mutual inclination $I$ ({\it top}), the eccentricities $e_\p$ (blue) and $e_\c$ (green) ({\it middle}), and the semi-major axis $a_\p$ ({\it bottom}).   \llabel{fig12}}
\end{figure}


\subsection{Effect of the dissipation rate}

\begin{figure}[ht!]
\includegraphics[width=8.5cm]{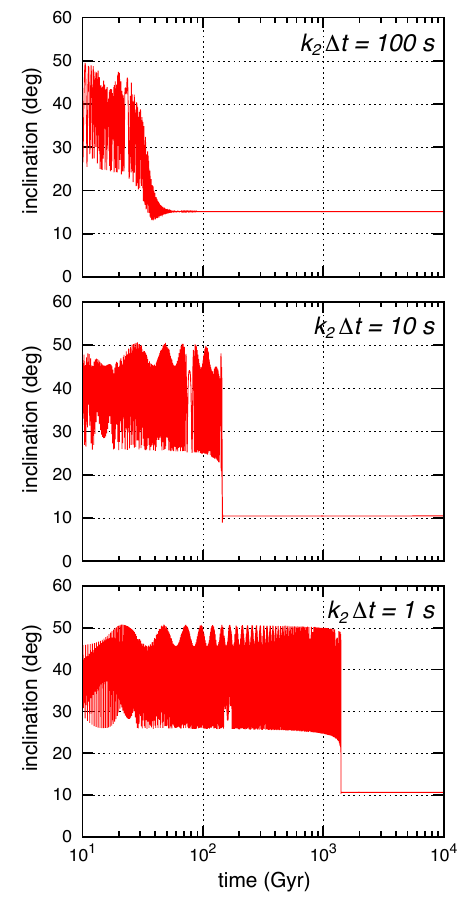}
\caption{Possible past evolution of the HD\,74156 system with $I_0 = 50^\circ$, using different dissipation rates. We plot the mutual inclination $I$ for $k_2 \Delta t = 100$\,s ({\it top}), $k_2 \Delta t = 10$\,s ({\it middle}), and $k_2 \Delta t = 1$\,s ({\it bottom}).   \llabel{figN}}
\end{figure}

In the former sections we have been using $k_2 \Delta t = 100 $\,s, or, in terms of $Q$-factor, $ k_2 / Q_\pp \approx 1.4 \times 10^{-4}$ (Eq.\,\ref{120704a}), which is similar to the present value measured for Saturn $ k_2 / Q_\pp = (2.3 \pm 0.7) \times 10^{-4}$ \citep[][]{Lainey_etal_2012}.
We adopted this value mainly for a better comparison with the previous paper on the planar case \citep{Correia_etal_2012}.
Other works on the tidal evolution of hot-Jupiters also adopted similar values for $\Delta t$ \citep[e.g.][]{Fabrycky_Tremaine_2007, Correia_etal_2011}.

However, the $Q$-factor of planets is unknown, and may vary by some orders of magnitude \citep[e.g.][]{Goldreich_Soter_1966}.
Indeed, the value measured for Jupiter appears to differ by a factor of ten, $ k_2 / Q_\pp = (1.1 \pm 0.2) \times 10^{-5}$ \citep[][]{Lainey_etal_2009}.
On the other hand, statistical studies on the observed eccentricity distribution of hot-Jupiters give $k_2 / Q_\pp \sim 10^{-6}$ \citep[e.g.][]{Jackson_etal_2008a, Hansen_2010}.
Part of the problem is that the nature of tidal dissipation in these planets is still poorly understood. 
In addition, the $Q$-factor is frequency dependent (i.e., model dependent), and therefore sometimes it is difficult to translate from one system to another.
In order to test the robustness of the inclination damping, here we test our model with lower dissipation rates (higher $Q$ values).

According to the tidal equations (Eqs.\,\ref{090515a}$-$\ref{120529a}), the evolution timescales are linearly proportional to $Q_\pp$, so higher $Q$-values delay the final evolution of a system.
In Figure~\ref{figN} we plot the evolution of the HD\,74156 system for three different dissipation values, $k_2 \Delta t_\pp = 1,\, 10, \, 100$\,s (which is equivalent to $k_2/Q \approx 10^{-6}, \, 10^{-5}, \, 10^{-4}$, respectively), starting with $I_0 = 50^\circ$. 
We observe that, although the evolution timescale is longer for higher $Q_\pp$ values (as expected), the inclination damping is still present.

In section~\ref{constraints} we saw that when we increase the initial mutual inclination $I_0$, the evolution timescale decreases very fast.
For instance, in the case of the HD\,74156 system, for $I_0 = 40^\circ$ the inner orbit becomes fully damped after 1\,Gyr, for  $I_0 = 50^\circ$ it takes about 100\,Myr, and for $I_0 = 60^\circ$ only 10\,Myr.
Therefore, for different dissipation rates we can still put constraints on the mutual inclination of hierarchical systems, the only consequence is that as we increase $Q_\pp$, the maximal mutual inclination that one can expect to observe also increases.

\section{Conclusion}

Many two-planet systems have been reported in hierarchical configurations.
For most of these systems the mutual inclinations are unknown. However, since the orbital eccentricities are typically high, we may expect that the formation mechanism that increased the eccentricities also increased the mutual inclinations.
Very often the innermost planet in these systems is close enough to the star to undergo tidal dissipation, which can pose constraints on the final evolution.

Here we have studied a particular subgroup of hierarchical planetary systems in which the inner planet's semi-major axis $0.1 < a_\p < 0.5$~AU (we called these ``moderate close-in planets'').
This range is important to ensure that the spin of the inner planet is fully evolved, but not its orbit.
Using an averaged secular model that takes into account gravitational interactions
up to  octupole order, we showed that for many initial conditions the mutual inclination is damped to relatively low values ($ I \sim 15^\circ $ for HD\,74156) on timescales shorter than the age of the system (less than one Gyr).

Without planetary perturbations and for zero obliquity there is no effect from tides on the evolution of the inclination (Eq.\,\ref{120529a}).
The inclination damping is thus not a direct consequence of tidal effects on the orbits.
%
%
The key element is a inner planet's eccentricity oscillation at a secular timescale similar to the synchronization time of its spin.
Indeed, the rotation (and thus the flattening (Eq.\,\ref{101220a})) of the planet is driven by its eccentricity variations (Eq.\,\ref{090515a}). 
In response to these excitations, the rotation is phase-shifted (Eq.\,\ref{110920a}) and the lag 
tends to pump the eccentricity (Eq.\,\ref{110902b}). 
%
%
%
Because the total angular momentum must be conserved, the increase in the inner orbit's eccentricity is accompanied by a subsequent reduction of the mutual inclination.
When the eccentricity pumping ceases, the inclination damping also stops.

For high mutual inclination values, quadrupolar gravitational exchanges with the eccentricity are more efficient, and so is the eccentricity pumping.
As a consequence, the inner orbit's eccentricity reaches higher values, tidal effects are enhanced at the periastron, and the system evolves on shorter timescales.
A strong inclination damping is then often followed by a fast circularization of the inner orbit.
Since most of the observed hierarchical systems still present substantial values of the inner-orbit's eccentricity after several Gyr, we expect that they cannot have very high mutual inclinations.
In particular, we are able to set constraints on the highest mutual inclination in these systems.

The evolution timescale also depends on the argument of the periastron in the invariant plane of the system, which is unknown at present for most systems.
Indeed, the uncertainty on this parameter is much higher than the uncertainty on the dissipation time lag $\Delta t$.
If $ \w_\p $ is in circulation, the evolution can be extremely fast (a few million years) for values close to $ \w_\p = 90^\circ $ or  $ \w_\p = 270^\circ $.
However, for high mutual inclinations $ \w_\p $ can be in Lidoz-Kozai libration for values close to $ \w_\p = 90^\circ $ or  $ \w_\p = 270^\circ $.
In that case, the evolution timescale can be delayed to several Gyr, but the equilibrium is unstable and broken when the system crosses the separatrix with the circulation regime.
After that, the inner planet is most likely lost.

\appendix

\section{Inclination damping}

\llabel{InclinationDamping}

According to expression (\ref{120529a}), the direct contribution of tidal effects to the inclination can be neglected for low obliquities.
We adopted $\varepsilon = 0^\circ$ for simplicity, but in a more realistic situation the obliquity is expected to be trapped in a Cassini state \citep[e.g.][]{Correia_etal_2011}.
Although high-obliquity states are possible, they are unlikely for close-in planets \citep{Levrard_etal_2007}, so we only expect these planets to be trapped in low-obliquity Cassini states, and hence $ d I / d t  \approx 0 $.

When tidal effects are combined with planetary perturbations, some counter-intuitive behavior may appear \citep[e.g.][]{Wu_Murray_2003, Mardling_2007, Correia_etal_2012}.
In particular, it has been shown that for moderate close-in planets, the gravitational perturbations of a distant coplanar companion combined with tides can increase the eccentricity of the inner orbit to very high values \citep{Correia_etal_2012}.
Because of the total angular momentum conservation, the eccentricity of the outer orbit is simultaneously decreased.

For coplanar systems ($ \sin I = 0 $), the octupolar terms in the expansion of the potential (Eq.\,\ref{090514a}) are the main responsible for the eccentricity variations of the inner orbit (Eq.\,\ref{110816h}).
However, for mutually inclined systems, the quadrupolar terms become dominating, and we can neglect the octupole contribution:
\be
\dot e_\p \approx \nu_{21} \frac{ \frac{5}{2}  (1-e_\p^2)^{1/2}  \sin^2 I}{(1-e_\c^2)^{3/2}} e_\p \sin 2 \w_\p \ . \llabel{120621a}
\ee
In addition, the contributions to the outer orbit eccentricity vanish ($\dot e_\c \approx 0$), meaning that the angular momentum exchanges only occur between $e_\p$ and $I$ (Eq.\,\ref{120329d}).
Thus, if somehow the quadrupole interactions are able to pump the inner orbit eccentricity as in the planar case \citep{Correia_etal_2012}, one can expect to observe a decrease in the mutual inclination of the system.

To understand this mechanism, we can simplify equations of motion without loss of generality.
Since we assume that the orbits are not coplanar, we retain only the quadrupolar terms for the gravitational perturbations, that is, the conservative motion can be described solely by expression (\ref{120621a}) and the first four terms in expression (\ref{110819a1}):
\begin{eqnarray}
\dot \w_\p &\approx& \frac{\nu_\gr}{ (1-e_\p^2)} + \frac{\nu_1 \, x_\pp^2}{(1-e_\p^2)^2} \crm 
&+& \nu_{21}  \frac{2 (1-e_\p^2) + \frac{5}{2} (e_\p^2 - \sin^2 I) (1 - \cos 2 \w_\p)}{(1-e_\p^2)^{1/2}  (1-e_\c^2)^{3/2}} \crm 
&+& \nu_{22} \frac{(1 + \frac{3}{2} e_\p^2 - \frac{5}{2} e_\p^2 \cos 2 \w_\p ) \cos I}{(1-e_\c^2)^{2}} 
\ . \llabel{120621b}
\end{eqnarray}

We can neglect tidal effects on orbital quantities  (Eqs.\,\ref{090515b},\,\ref{090515c}), which is justified since $ \Kl_\pp\ll K_\pp $ (Eq.\ref{090514m}).
The only contribution of tides is then to the rotation rate (Eq.\ref{090515a}).
The semi-major axis and the mean motion are thus constant, as is the eccentricity of the outer orbit and $G_\c$.

For simplicity, we average expression (\ref{120621b}) over $ \w_\p$, and linearize the set of equations of motion in the vicinity of the averaged values of $x$, $e_\p$, and $I$.
Let $x=x_0+\delta x$, where $x_0$ is the solution of (\ref{090520a}), $e_\p = e_{\p 0} +
\delta e_\p$, and $I = I_{0} + \delta I$. 
In the following,
$\delta I$ is expressed as a function of  $\delta e_\p$
using the angular momentum conservation (Eq.\ref{120329d}):
\be
\delta I  \approx - \frac{1}{\sin I_0} \left( \frac{G_\p}{G_\c} + \cos I_0 \right) \frac{e_{\p 0}}{1-e_{\p 0}^2} \delta e_\p \ . \llabel{120621c}
\ee

Then, the set of equations of motion (\ref{120621a}, \ref{120621b}, and \ref{090515a}) reduces to
\be
\delta \dot e_\p = A \sin 2 \w_\p \ , \llabel{110812b}
\ee
\be
\dot \w_\p = g + g_x \delta x + g_e \delta e_\p\ , \llabel{110812c}
\ee
\be
\delta \dot x = -\nu_x \delta x + \nu_e \delta e_\p \ , \llabel{110812d}
\ee
with 
\be 
A = \nu_{21} \frac{ \frac{5}{2}  (1-e_{\p 0}^2)^{1/2}  \sin^2 I_0}{(1-e_\c^2)^{3/2}} e_{\p 0}
\ , \llabel{121217c}
\ee
\begin{eqnarray} 
g &=&\frac{\nu_\gr}{ (1-e_{\p 0}^2)} +  \nu_{21}  \frac{2 (1-e_{\p 0}^2) + \frac{5}{2} (e_{\p 0}^2 - \sin^2 I_0)}{(1-e_{\p 0}^2)^{1/2}  (1-e_\c^2)^{3/2}} \crm 
&& + \frac{\nu_1 \, x_{\pp 0}^2}{(1-e_{\p 0}^2)^2} + \nu_{22} \frac{(1 + \frac{3}{2} e_{\p 0}^2) \cos I_0}{(1-e_\c^2)^{2}}  \ , \llabel{121217d}
\end{eqnarray}
\be g_x = \nu_1 \frac{2 x_{\pp 0}}{(1-e_{\p 0}^2)^2} 
\ ,
\ee
\begin{eqnarray}
g_e &=& \nu_0 \frac{2 e_{\p 0}}{(1-e_{\p 0}^2)^2} + \nu_1 \frac{4 x_0^2 e_{\p
0}}{(1-e_{\p 0}^2)^3} \crm 
&& + \nu_{21} \frac{10 \cos^2 I_0 e_{\p 0}}{(1-e_{\p 0}^2)^2 (1-e_\c^2)^{3/2}} \crm 
&& + \nu_{21} \frac{(1+ \frac{3}{2} e_{\p 0}^2) e_{\p 0}}{(1-e_{\p 0}^2)^{1/2}  (1-e_\c^2)^{5/2}} \crm
&& + \nu_{22} \frac{(4 - \frac{3}{2} e_{\p 0}^2) e_{\p 0} \cos I_0}{(1-e_{\p 0}^2) (1-e_\c^2)^2}  \crm
&& + \nu_{22} \frac{5 e_{\p 0} \cos I_0}{(1-e_{\p 0}^2)^{3/2} (1-e_\c^2)^2}
\ , 
\end{eqnarray}
\be 
\nu_x =  K_\pp f_1(e_{\p 0}) \ , \llabel{121217a}
\ee
\be 
\nu_e = -K_\pp ( f'_1 (e_{\p 0}) x_0 - f'_2(e_{\p 0}) ) \ ,  \llabel{121217b}
\ee
where
$f'_1(e) = 15 (e+3 e^3/2 +e^5/8) / (1-e^2)^{11/2} $, and
$f'_2(e) =  3 (9e+65e^3/2+125e^5/8+5e^7/8) / (1-e^2)^{7} $.

At first order, the precession of the periastron is constant
$ \dot \w_\p \simeq g $, 
and the eccentricity is simply given from expression (\ref{110812b}) as
\be
\delta e_\p = - \Deltae \cos  (2 g t + 2 \w_{\p 0}) \ , \llabel{110812e}
\ee
where $ \Deltae = A / 2 g $, and $ \w_\p =  g t + \w_{\p 0} $. 
That is, the eccentricity $e_\p$ presents periodic variations around an equilibrium value $ e_{ \p 0} $, with amplitude $ \Deltae $ and frequency $ 2 g $.
Since $ g_x \delta x, g_e \delta e_\p \ll g$, the above solution for the eccentricity can be adopted as the zeroth-order solution of the system of equations (\ref{110812b}$-$\ref{110812d}).
With this approximation, the equation of motion of $\delta x$ (\ref{110812d}) becomes that of a driven harmonic oscillator whose steady state solution is
\be
\delta x = - \Deltax \cos ( 2 g t + 2 \w_{\p 0} - \phi) \ , \llabel{110920a}
\ee
with $ \Deltax = \nu_e \Deltae / \sqrt{\nu_x^2 + 4 g^2 } $, 
and $\sin \phi = 2 g / \sqrt{\nu_x^2 + 4 g^2 } $. The rotation rate 
thus presents an oscillation identical to the eccentricity (Eq.\ref{110812e}), but with smaller amplitude and delayed by an angle $\phi$
\citep[see][]{Correia_2011}.
Using the above expression in equation (\ref{110812c}) and integrating, gives for the periastron
\be
\w_\p = g t + \w_{\p 0} - \frac{g_x \Deltax}{2 g}  \sin (2 g t + 2 \w_{\p 0} - \phi)
- \frac{g_e \Deltae}{2 g}  \sin (2 g t + \w_{\p 0})
\llabel{110812g} \ .
\ee
Finally, substituting in expression (\ref{110812b}) and using the approximation $ g_x \Deltax, g_e \Deltae \ll g$ gives
\begin{eqnarray}
\delta \dot e_\p 
& \approx & A \sin (2 g t + 2 \w_{\p 0}) \crm
& & - \frac{g_e \Deltae}{2 g} A  \sin (2 g t + 2 \w_{\p 0}) \cos  (2 g t + 2 \w_{\p 0}) \crm 
& & - \frac{g_x \Deltax}{2 g} A \sin (2 g t + 2 \w_{\p 0} - \phi) \cos  (2 g t + 2 \w_{\p 0}) 
  \llabel{110812h} \ ,
\end{eqnarray}
or, combining the two products of periodic functions,
\begin{eqnarray}
\delta \dot e_\p  & = & A \sin (2 g t +  2 \w_{\p 0}) - \frac{g_e \Deltae}{4 g} A \sin (4 g t + 4 \w_{\p 0})
 \crm
& &  - \frac{g_x \Deltax}{4 g}  A \sin (4 g t + 4 \w_{\p 0} - \phi)  +  \frac{g_x A}{4 g} \Deltax \sin \phi
\llabel{110812i} \ .
\end{eqnarray}

The two middle terms in the above equation can be neglected since they are
periodic and have a very small amplitude ($ g_x \Deltax, g_e \Deltae \ll g $). However, the last term in $\sin \phi$ is constant and it adds a small drift to the eccentricity,
\be
\overline {\delta \dot e_\p} =  \frac{g_x A}{4 g} \Deltax \sin \phi = \frac{\nu_e g_x A^2}{4 g (\nu_x^2 + 4 g^2 )}  \ . \llabel{110902b}
\ee
Note that the phase lag $\phi$ between the eccentricity (Eq.\,\ref{110812e}) and the rotation variations (Eq.\,\ref{110920a}) is essential to obtain a drift on the eccentricity.
From expressions (\ref{121217c}) and (\ref{121217d}) we have $A \sim \nu_{21} \sim g$, while from expressions (\ref{121217a}) and (\ref{121217b}) we get $\nu_x \sim \nu_e \sim K_\pp$.
Consequently,
\be
\sin \phi = \frac{2 g}{\sqrt{\nu_x^2 + 4 g^2}} \, \sim 2 \frac{g}{\sqrt{K_\pp^2 + 4 g^2}} \ ,
\ee
and
\be
\Delta x = \frac{\nu_e A}{2 g \sqrt{\nu_x^2 + 4 g^2}} \, \sim \frac{K_\pp}{2 \sqrt{K_\pp^2 + 4 g^2}} \, \sim \frac{1}{2} \cos \phi\ ,
\ee
that is,
\be
\overline {\delta \dot e_\p} \sim \frac{g_x}{16} \sin 2\phi \, \sim \frac{\nu_1}{8} \sin 2 \phi \ .
\llabel{121217e}
\ee
Thus, the drift vanishes when $\phi=0$ or $90$ degrees, i.e., for strong dissipation ($K_\pp \gg g $), where $\overline {\delta \dot e_\p} \sim \nu_1 g / K$, or for weak dissipation ($ K_\pp \ll g $), where $\overline {\delta \dot e_\p} \sim \nu_1 K / g $, respectively.
The pumping effect on the eccentricity is then maximized when $\phi = 45^\circ$, for $  K_\pp \sim g $, with $\overline {\delta \dot e_\p} \sim \nu_1 $.


We therefore conclude that the quadrupolar perturbation of an inclined companion enhances the effect of the octupolar terms in the planar case \citep{Correia_etal_2012}.
However, since the angular momentum is exchanged with the inclination, by assuming $G_\c \gg G_\p $ we obtain from expression (\ref{120621c}) that $ \overline {\delta I} < 0 $ for $ I_0 < 90^\circ$, and $ \overline {\delta I} > 0 $ for $ I_0 > 90^\circ$. That is, the mutual inclination is damped as long as the inner orbit eccentricity is pumped.

The main difference when we consider the full non-linearized problem is that the drift in the eccentricity (Eq.\,\ref{110902b}) cannot grow indefinitely.
Indeed, when the eccentricity reaches high values, the drift vanishes (Fig.\,\ref{fig5}b).
Moreover, tidal effects are also enhanced for high eccentricities and counterbalance the drift (Eq.\,\ref{090515c}).
Hence, the drift in the eccentricity is never permanent, neither is the inclination damping, although they can last for the entire age of the system.

\begin{acknowledgements}
We wish to thank the referee Rosemary Mardling, who made very valuable suggestions.
We acknowledge support by PNP-CNRS, CS of Paris Observatory,
PICS05998 France-Portugal program, the European Research Council/European
Community under the FP7 through a Starting Grant, and Funda\c{c}\~ao
para a Ci\^encia e a Tecnologia, Portugal (grants PTDC/CTE-AST/098528/2008, SFRH/BSAB/1148/2011, PEst-C/CTM/LA0025/2011).
\end{acknowledgements}

\bibliographystyle{aa}
\bibliography{correia}

\begin{thebibliography}{51}
\expandafter\ifx\csname natexlab\endcsname\relax\def\natexlab#1{#1}\fi

\bibitem[{{Bean} {et~al.}(2008){Bean}, {McArthur}, {Benedict}, \&
  {Armstrong}}]{Bean_etal_2008}
{Bean}, J.~L., {McArthur}, B.~E., {Benedict}, G.~F., \& {Armstrong}, A. 2008,
  \apj, 672, 1202

\bibitem[{{Benedict} {et~al.}(2010){Benedict}, {McArthur}, {Bean}, {Barnes},
  {Harrison}, {Hatzes}, {Martioli}, \& {Nelan}}]{Benedict_etal_2010}
{Benedict}, G.~F., {McArthur}, B.~E., {Bean}, J.~L., {et~al.} 2010, \aj, 139,
  1844

\bibitem[{{Butler} {et~al.}(2006){Butler}, {Wright}, {Marcy}, {Fischer},
  {Vogt}, {Tinney}, {Jones}, {Carter}, {Johnson}, {McCarthy}, \&
  {Penny}}]{Butler_etal_2006}
{Butler}, R.~P., {Wright}, J.~T., {Marcy}, G.~W., {et~al.} 2006, \apj, 646, 505

\bibitem[{{Chatterjee} {et~al.}(2008){Chatterjee}, {Ford}, {Matsumura}, \&
  {Rasio}}]{Chatterjee_etal_2008}
{Chatterjee}, S., {Ford}, E.~B., {Matsumura}, S., \& {Rasio}, F.~A. 2008, \apj,
  686, 580

\bibitem[{{Correia}(2009)}]{Correia_2009}
{Correia}, A.~C.~M. 2009, \apjl, 704, L1

\bibitem[{{Correia}(2011)}]{Correia_2011}
{Correia}, A.~C.~M. 2011, in IAU Symposium, Vol. 276, IAU Symposium, ed.
  {A.~Sozzetti, M.~G.~Lattanzi, \& A.~P.~Boss}, 287--294

\bibitem[{{Correia} {et~al.}(2012){Correia}, {Bou{\'e}}, \&
  {Laskar}}]{Correia_etal_2012}
{Correia}, A.~C.~M., {Bou{\'e}}, G., \& {Laskar}, J. 2012, \apjl, 744, L23

\bibitem[{{Correia} \& {Laskar}(2004)}]{Correia_Laskar_2004}
{Correia}, A.~C.~M. \& {Laskar}, J. 2004, \nat, 429, 848

\bibitem[{{Correia} \& {Laskar}(2010{\natexlab{a}})}]{Correia_Laskar_2010}
{Correia}, A.~C.~M. \& {Laskar}, J. 2010{\natexlab{a}}, Icarus, 205, 338

\bibitem[{{Correia} \& {Laskar}(2010{\natexlab{b}})}]{Correia_Laskar_2010B}
{Correia}, A.~C.~M. \& {Laskar}, J. 2010{\natexlab{b}}, in Exoplanets
  (University of Arizona Press), 534--575

\bibitem[{{Correia} \& {Laskar}(2012)}]{Correia_Laskar_2012}
{Correia}, A.~C.~M. \& {Laskar}, J. 2012, \apjl, 751, L43

\bibitem[{{Correia} {et~al.}(2011){Correia}, {Laskar}, {Farago}, \&
  {Bou{\'e}}}]{Correia_etal_2011}
{Correia}, A.~C.~M., {Laskar}, J., {Farago}, F., \& {Bou{\'e}}, G. 2011,
  Celestial Mechanics and Dynamical Astronomy, 111, 105

\bibitem[{{Correia} {et~al.}(2003){Correia}, {Laskar}, \& {N\'eron de
  Surgy}}]{Correia_etal_2003}
{Correia}, A.~C.~M., {Laskar}, J., \& {N\'eron de Surgy}, O. 2003, Icarus, 163,
  1

\bibitem[{{Correia} {et~al.}(2005){Correia}, {Udry}, {Mayor}, {Laskar}, {Naef},
  {Pepe}, {Queloz}, \& {Santos}}]{Correia_etal_2005}
{Correia}, A.~C.~M., {Udry}, S., {Mayor}, M., {et~al.} 2005, \aap, 440, 751

\bibitem[{{Couetdic} {et~al.}(2010){Couetdic}, {Laskar}, {Correia}, {Mayor}, \&
  {Udry}}]{Couetdic_etal_2010}
{Couetdic}, J., {Laskar}, J., {Correia}, A.~C.~M., {Mayor}, M., \& {Udry}, S.
  2010, \aap, 519, A10

\bibitem[{{Efroimsky} \& {Williams}(2009)}]{Efroimsky_Williams_2009}
{Efroimsky}, M. \& {Williams}, J.~G. 2009, Celestial Mechanics and Dynamical
  Astronomy, 104, 257

\bibitem[{{Fabrycky} \& {Tremaine}(2007)}]{Fabrycky_Tremaine_2007}
{Fabrycky}, D. \& {Tremaine}, S. 2007, \apj, 669, 1298

\bibitem[{{Ford} {et~al.}(2000){Ford}, {Kozinsky}, \& {Rasio}}]{Ford_etal_2000}
{Ford}, E.~B., {Kozinsky}, B., \& {Rasio}, F.~A. 2000, \apj, 535, 385

\bibitem[{{Giguere} {et~al.}(2012){Giguere}, {Fischer}, {Howard}, {Johnson},
  {Henry}, {Wright}, {Marcy}, {Isaacson}, {Hou}, \&
  {Spronck}}]{Giguere_etal_2012}
{Giguere}, M.~J., {Fischer}, D.~A., {Howard}, A.~W., {et~al.} 2012, \apj, 744,
  4

\bibitem[{{Giuppone} {et~al.}(2012){Giuppone}, {Morais}, {Bou{\'e}}, \&
  {Correia}}]{Giuppone_etal_2012}
{Giuppone}, C.~A., {Morais}, M.~H.~M., {Bou{\'e}}, G., \& {Correia}, A.~C.~M.
  2012, \aap, 541, A151

\bibitem[{{Goldreich} \& {Soter}(1966)}]{Goldreich_Soter_1966}
{Goldreich}, P. \& {Soter}, S. 1966, Icarus, 5, 375

\bibitem[{{Hansen}(2010)}]{Hansen_2010}
{Hansen}, B.~M.~S. 2010, \apj, 723, 285

\bibitem[{{Jackson} {et~al.}(2008){Jackson}, {Greenberg}, \&
  {Barnes}}]{Jackson_etal_2008a}
{Jackson}, B., {Greenberg}, R., \& {Barnes}, R. 2008, \apj, 678, 1396

\bibitem[{{Kozai}(1962)}]{Kozai_1962}
{Kozai}, Y. 1962, \aj, 67, 591

\bibitem[{{Lainey} {et~al.}(2009){Lainey}, {Arlot}, {Karatekin}, \& {van
  Hoolst}}]{Lainey_etal_2009}
{Lainey}, V., {Arlot}, J.-E., {Karatekin}, {\"O}., \& {van Hoolst}, T. 2009,
  Nature, 459, 957

\bibitem[{{Lainey} {et~al.}(2012){Lainey}, {Karatekin}, {Desmars}, {Charnoz},
  {Arlot}, {Emelyanov}, {Le Poncin-Lafitte}, {Mathis}, {Remus}, {Tobie}, \&
  {Zahn}}]{Lainey_etal_2012}
{Lainey}, V., {Karatekin}, {\"O}., {Desmars}, J., {et~al.} 2012, \apj, 752, 14

\bibitem[{{Lambeck}(1988)}]{Lambeck_1988}
{Lambeck}, K. 1988, {Geophysical geodesy : the slow deformations of the earth
  Lambeck.} (Oxford [England] : Clarendon Press ; New York : Oxford University
  Press, 1988.)

\bibitem[{{Laskar}(1990)}]{Laskar_1990}
{Laskar}, J. 1990, Icarus, 88, 266

\bibitem[{{Laskar}(1993)}]{Laskar_1993PD}
{Laskar}, J. 1993, Physica D Nonlinear Phenomena, 67, 257

\bibitem[{{Laskar} \& {Bou{\'e}}(2010)}]{Laskar_Boue_2010}
{Laskar}, J. \& {Bou{\'e}}, G. 2010, \aap, 522, A60

\bibitem[{{Laskar} {et~al.}(2012){Laskar}, {Bou{\'e}}, \&
  {Correia}}]{Laskar_etal_2012}
{Laskar}, J., {Bou{\'e}}, G., \& {Correia}, A.~C.~M. 2012, \aap, 538, A105

\bibitem[{{Lee} \& {Peale}(2003)}]{Lee_Peale_2003}
{Lee}, M.~H. \& {Peale}, S.~J. 2003, \apj, 592, 1201

\bibitem[{{Levrard} {et~al.}(2007){Levrard}, {Correia}, {Chabrier}, {Baraffe},
  {Selsis}, \& {Laskar}}]{Levrard_etal_2007}
{Levrard}, B., {Correia}, A.~C.~M., {Chabrier}, G., {et~al.} 2007, \aap, 462,
  L5

\bibitem[{{Lidov}(1961)}]{Lidov_1961}
{Lidov}, M.~L. 1961, Iskus. sputniky Zemly (in Russian), 8, 5

\bibitem[{{Lidov}(1962)}]{Lidov_1962}
{Lidov}, M.~L. 1962, \planss, 9, 719

\bibitem[{{Marchal}(1990)}]{Marchal_1990}
{Marchal}, C. 1990, {The Three-Body Problem} (Elsevier, Amsterdam)

\bibitem[{{Mardling}(2007)}]{Mardling_2007}
{Mardling}, R.~A. 2007, \mnras, 382, 1768

\bibitem[{{Meschiari} {et~al.}(2011){Meschiari}, {Laughlin}, {Vogt}, {Butler},
  {Rivera}, {Haghighipour}, \& {Jalowiczor}}]{Meschiari_etal_2011}
{Meschiari}, S., {Laughlin}, G., {Vogt}, S.~S., {et~al.} 2011, \apj, 727, 117

\bibitem[{{Mignard}(1979)}]{Mignard_1979}
{Mignard}, F. 1979, Moon and Planets, 20, 301

\bibitem[{{Murray} \& {Dermott}(1999)}]{Murray_Dermott_1999}
{Murray}, C.~D. \& {Dermott}, S.~F. 1999, {Solar System Dynamics} (Cambridge
  University Press)

\bibitem[{{Naef} {et~al.}(2004){Naef}, {Mayor}, {Beuzit}, {Perrier}, {Queloz},
  {Sivan}, \& {Udry}}]{Naef_etal_2004}
{Naef}, D., {Mayor}, M., {Beuzit}, J.~L., {et~al.} 2004, \aap, 414, 351

\bibitem[{{Nagasawa} {et~al.}(2008){Nagasawa}, {Ida}, \&
  {Bessho}}]{Nagasawa_etal_2008}
{Nagasawa}, M., {Ida}, S., \& {Bessho}, T. 2008, \apj, 678, 498

\bibitem[{{Pilyavsky} {et~al.}(2011){Pilyavsky}, {Mahadevan}, {Kane}, {Howard},
  {Ciardi}, {de Pree}, {Dragomir}, {Fischer}, {Henry}, {Jensen}, {Laughlin},
  {Marlowe}, {Rabus}, {von Braun}, {Wright}, \& {Wang}}]{Pilyavsky_etal_2011}
{Pilyavsky}, G., {Mahadevan}, S., {Kane}, S.~R., {et~al.} 2011, \apj, 743, 162

\bibitem[{{Pont} {et~al.}(2011){Pont}, {Husnoo}, {Mazeh}, \&
  {Fabrycky}}]{Pont_etal_2011}
{Pont}, F., {Husnoo}, N., {Mazeh}, T., \& {Fabrycky}, D. 2011, \mnras, 414,
  1278

\bibitem[{{S{\'e}gransan} {et~al.}(2010){S{\'e}gransan}, {Udry}, {Mayor},
  {Naef}, {Pepe}, {Queloz}, {Santos}, {Demory}, {Figueira}, {Gillon},
  {Marmier}, {M{\'e}gevand}, {Sosnowska}, {Tamuz}, \&
  {Triaud}}]{Segransan_etal_2010}
{S{\'e}gransan}, D., {Udry}, S., {Mayor}, M., {et~al.} 2010, \aap, 511, A45

\bibitem[{{Singer}(1968)}]{Singer_1968}
{Singer}, S.~F. 1968, \gjras, 15, 205

\bibitem[{{Touma} {et~al.}(2009){Touma}, {Tremaine}, \&
  {Kazandjian}}]{Touma_etal_2009}
{Touma}, J.~R., {Tremaine}, S., \& {Kazandjian}, M.~V. 2009, \mnras, 394, 1085

\bibitem[{{Vogt} {et~al.}(2005){Vogt}, {Butler}, {Marcy}, {Fischer}, {Henry},
  {Laughlin}, {Wright}, \& {Johnson}}]{Vogt_etal_2005}
{Vogt}, S.~S., {Butler}, R.~P., {Marcy}, G.~W., {et~al.} 2005, \apj, 632, 638

\bibitem[{{Wright} {et~al.}(2009){Wright}, {Upadhyay}, {Marcy}, {Fischer},
  {Ford}, \& {Johnson}}]{Wright_etal_2009}
{Wright}, J.~T., {Upadhyay}, S., {Marcy}, G.~W., {et~al.} 2009, \apj, 693, 1084

\bibitem[{{Wu} \& {Goldreich}(2002)}]{Wu_Goldreich_2002}
{Wu}, Y. \& {Goldreich}, P. 2002, \apj, 564, 1024

\bibitem[{{Wu} \& {Murray}(2003)}]{Wu_Murray_2003}
{Wu}, Y. \& {Murray}, N. 2003, \apj, 589, 605

\end{thebibliography}

\end{document}